\documentclass[fleqn,10pt]{olplainarticle}

\usepackage{amsmath, amssymb, amsthm}
\usepackage[colorlinks]{hyperref}

\def\tmp#1#2#3{%
  \definecolor{Hy#1color}{#2}{#3}%
  \hypersetup{#1color=Hy#1color}}
\tmp{link}{HTML}{27AB4F}
\tmp{cite}{HTML}{659DBD}
\tmp{file}{HTML}{131877}
\tmp{url} {HTML}{8A0087}
\tmp{menu}{HTML}{727500}
\tmp{run} {HTML}{137776}
\def\tmp#1#2{%
  \colorlet{Hy#1bordercolor}{Hy#1color#2}%
  \hypersetup{#1bordercolor=Hy#1bordercolor}}
\tmp{link}{!60!white}
\tmp{cite}{!60!white}
\tmp{file}{!60!white}
\tmp{url} {!60!white}
\tmp{menu}{!60!white}
\tmp{run} {!60!white}

\usepackage{csquotes}
\usepackage{url}
\usepackage{stix}
\usepackage[T1]{fontenc}

\usepackage[acronym, smallcaps]{glossaries}
\newacronym{rwm}{rwm}{random walk Metropolis}
\newacronym{esjd}{esjd}{expected squared jumping distance}
\newacronym{pdf}{pdf}{probability density function}
\newacronym{mcmc}{mcmc}{Markov chain Monte Carlo}
\newacronym{iid}{iid}{independent and identically distributed}
\newacronym{ess}{ess}{effective sample size}
\newacronym{rhs}{rhs}{right-hand side}
\glsdisablehyper

\newcommand{\btheta}{\boldsymbol{\theta}}
\newcommand{\bz}{\mathbf{z}}
\renewcommand{\d}{\mathrm{d}}
 \DeclareMathOperator{\R}{\mathbb{R}}
 \newcommand{\E}{\mathbb{E}}
  \newcommand{\Prob}{\mathbb{P}}

\newtheorem{theorem}{Theorem}
\newtheorem{proposition}{Proposition}
\newtheorem{assumption}{Assumption}
\newtheorem{corollary}{Corollary}

\newcommand{\F}[1]{\mathcal{I}{(#1)}}
\newcommand{\iF}[1]{\F{#1}^{-1}}

\title{Optimal scaling of random-walk Metropolis algorithms using Bayesian large-sample asymptotics}

\author[1, *]{Sebastian M Schmon}
\author[2, *]{Philippe Gagnon}
\affil[1]{Improbable and University of Durham}
\affil[2]{Université de Montréal}
\affil[*]{equal contribution}


\begin{abstract}
High-dimensional limit theorems have been shown useful to derive tuning rules for finding the optimal scaling in random-walk Metropolis algorithms. The assumptions under which weak convergence results are proved are however restrictive: the target density is typically assumed to be of a product form. Users may thus doubt the validity of such tuning rules in practical applications. In this paper, we shed some light on optimal-scaling problems from a different perspective, namely a large-sample one. This allows to prove weak convergence results under realistic assumptions and to propose novel parameter-dimension-dependent tuning guidelines. The proposed guidelines are consistent with previous ones when the target density is close to having a product form, and the results highlight that the correlation structure has to be accounted for to avoid performance deterioration if that is not the case, while justifying the use of a natural (asymptotically exact) approximation to the correlation matrix that can be employed for the very first algorithm run.
\end{abstract}

\begin{document}

\def\sectionautorefname{Section}
\def\subsectionautorefname{Section}
\def\corollaryautorefname{Corollary}
\def\propositionautorefname{Proposition}
\def\assumptionautorefname{Assumption}

\flushbottom
\maketitle
\thispagestyle{empty}

\section{Introduction}

\subsection{Random-walk Metropolis algorithms}

Consider a Bayesian statistical framework where one wants to sample from an intractable posterior distribution $\pi$ to perform inference. This posterior distribution, also called the \textit{target distribution} in a sampling context, is considered here to be that of model parameters $\boldsymbol\theta \in \boldsymbol\Theta = \R^d$, given a data sample of size $n$. We assume that $\pi$ has a \gls{pdf} with respect to the Lebesgue measure; to simplify, we will also use $\pi$ to denote this density function. Tools called \textit{\gls{rwm}} algorithms \citep{Metropolis1953}, which are \gls{mcmc} methods, can be employed to sample from $\pi$. An iteration of such an algorithm can be outlined as follows: given a current value of the chain $\boldsymbol\theta$, a proposal for the next one is made using
\begin{align*}
 \boldsymbol\theta' := \boldsymbol\theta + \mathbf{S} \, \boldsymbol\epsilon, \quad \boldsymbol\epsilon \sim \varphi(\,\cdot\, ; \mathbf{0}, \mathbf{1}),
\end{align*}
where $\mathbf{S}$ is a scaling matrix and $\varphi(\,\cdot\, ; \mathbf{0}, \mathbf{1})$ denotes the standard normal distribution; this proposal is accepted with probability
\[
 \alpha(\boldsymbol\theta, \boldsymbol\theta') := \min\left\{1, \frac{\pi(\boldsymbol\theta')}{\pi(\boldsymbol\theta)}\right\};
\]
if the proposal is rejected, the chain remains at the same state.

\subsection{Optimal-scaling problems}

Often, $\mathbf{S} = \lambda \mathbf{1}$, where $\lambda$ is a positive constant to be determined. In this case, $\lambda$ is the only free parameter. Yet, this parameter has to be tuned carefully because small values lead to tiny movements of the Markov chain simulated by \gls{rwm}, while large values induce high rejection rates, both being undesirable. Finding the optimal value is thus a non-trivial problem. The last twenty years have witnessed significant progress in the line of research studying such problems called \textit{optimal-scaling problems}, whether it is in \gls{rwm} \citep{RobertsGelmanGilks1997, bedard2007weak, sherlock2009optimal, durmus2017opimal, YANG20206094} or other algorithms including a scaling parameter \citep{Roberts1998, bedard2012scaling, beskos2013optimal}. In all these articles, the authors derive tuning rules based on analyses in the high-dimensional regime $d \rightarrow \infty$.

In the seminal work of \cite{RobertsGelmanGilks1997} on \gls{rwm}, the tuning rule for $\lambda$ follows from the analysis of a Langevin diffusion which is the limiting process of a re-scaled continuous-time version of \gls{rwm}.
The rule is remarkably simple: set $\lambda = \ell/\surd{d}$ and tune $\ell$ so that the acceptance rate is 0.234. The resulting optimal value is \textit{universal}, in the sense that it minimizes the stationary integrated autocorrelation time of \textit{any} function of the limiting process.
The tuning rule is, however, derived under the assumption that $\pi(\boldsymbol\theta) = \prod_{i = 1}^d f(\theta_i)$, where $\boldsymbol\theta := (\theta_1, \ldots, \theta_d)$ and $f$ satisfies some regularity conditions. Assuming \gls{iid} parameters considerably reduces the scope of applicability. One may be tempted to search for transformations/standardizations yielding \gls{iid} parameters to expand the scope, but they exist only in specific situations (e.g., Gaussian target distributions). It will be seen that one of the main contributions of this paper is to provide formal and realistic conditions under which \gls{rwm} targeting $\pi$ behave similarly to \gls{rwm} targeting a Gaussian distribution with specific mean and covariance in an asymptotic regime. Our results thus allow to demonstrate that standardizing the parameters to expand the scope of applicability of the results of \cite{RobertsGelmanGilks1997} is valid under regularity conditions, but only asymptotically.

The scope has been expanded otherwise in the past. For example, \cite{bedard2007weak} and \cite{durmus2017opimal} proved that the result is robust to departure from the \textit{identically distributed} part of the assumption.
\cite{YANG20206094} proved that the result is valid under assumptions that are more general but difficult to verify. Empirical results in realistic scenarios where the \gls{iid} assumption is, thus, not satisfied show that an acceptance rate of 0.234 is close to being optimal in these scenarios \citep[e.g.][]{shang2015monte, zhang2016power, gagnon2021PCR}, which can be seen as another demonstration of the robustness of the original results.

\subsection{Contributions}\label{sec:contributions}

In this paper, we provide an alternative explanation of these empirical results in realistic scenarios, based on Bayesian large-sample theory. To achieve this, we revisit optimal-scaling problems in \gls{rwm} by exploiting important results underpinning that theory. In particular, we prove a weak convergence result as $n \rightarrow \infty$, with $d$ being fixed, and derive tuning rules from it. While this asymptotic regime is ubiquitous in statistics, it is only recently that it was found useful in the analysis of \gls{mcmc} algorithms \citep{deligiannidis2015, gagnon2019RJ, Schmon2020}.
Intuitively, if $n$ is large enough and $\pi$ is a posterior distribution resulting from a sufficiently regular Bayesian model, then $\pi$ is close to a concentrating Gaussian, implying that \gls{rwm} algorithms targeting $\pi$ behave like those targeting a Gaussian. This idea is formalized in \autoref{sec:large_sample_asymptotic}.

The proximity between $\pi$ and a concentrating Gaussian can be established by virtue of Bernstein--von Mises theorems (see, e.g., Theorem 10.1 in \citealt{VanderVaart2000} and \citealt{kleijn2012}). Verifying that a Bayesian model is sufficiently regular is thus closely related to verifying that the assumptions of such theorems are satisfied and has a priori nothing to do with whether the parameters are \gls{iid} or not.
Instead, such theorems rely on \emph{local asymptotic normality}, meaning that a certain function of the $\log$-likelihood allows for a quadratic expansion (usually) around some \enquote{true} parameter value ${\boldsymbol\theta_0}$.
If the posterior concentrates around ${\boldsymbol\theta_0}$, the quadratic expansion of the $\log$-likelihood implies an asymptotically Gaussian posterior; this happens under weak conditions such as \gls{iid} \emph{data points} with regularity conditions on the distribution and positive prior mass around ${\boldsymbol\theta_0}$.
The results in \cite{RobertsGelmanGilks1997} actually rely on a similar quadratic expansion, but one that requires to impose a \gls{iid} constraint on the parameters instead. We discuss in more detail the resemblance between both expansions in \autoref{sec:analysis_and_tuning}, allowing to establish a connection between our guidelines and theirs.

An advantage of the approach adopted in this paper to analyse \gls{mcmc} algorithms is that a lot is known about which models are sufficiently regular \citep[e.g.][]{lecam1953some, bickel1969some, johnson1970asymptotic, ghosal1995convergence, VanderVaart2000, kleijn2012}. Many models based on the exponential family are, for instance, regular enough. A notable example of such a model, namely Bayesian logistic regression, is studied in \autoref{sec:logistic_regression}.

We finish this section by outlining our main contributions:
\begin{itemize}
\itemsep 0mm
    \item[(i)] presentation of a large-sample-asymptotic framework and realistic assumptions under which a weak convergence of \gls{rwm} is proved (\autoref{sec:large_sample_asymptotic});

	\item[(ii)] an extensive analysis of the limiting \gls{rwm} algorithm (\autoref{sec:analysis_and_tuning}) that allow to: (a) provide \emph{dimension-dependent} optimal tuning guidelines, (b) show
 that the \enquote{0.234} rule-of-thumb is asymptotically valid from the point of view adopted in this paper in certain situations and that this rule is in fact quite robust to a departure from the \gls{iid} assumption when $\mathbf{S} = \lambda \mathbf{1}$, without providing any guarantee regarding the algorithm performance; the latter deteriorates when there is a significant departure from the \gls{iid} assumption and $\mathbf{S} = \lambda \mathbf{1}$ because this scaling matrix does not account for the correlation in between the parameters;

	\item[(iii)] justification of the use of natural asymptotically exact approximations to the covariance matrix such as the inverse Fisher information or its observed version that can be employed for the very first algorithm run to avoid deterioration of performance (\autoref{sec:analysis_and_tuning}).
\end{itemize}

Our analysis is mainly based on an efficiency measure called the \textit{\gls{esjd}}. It is defined as the average squared distance between two consecutive states (or a function of them). Optimizing this measure does \textit{not} yield a universally optimal scaling because it is optimal for \textit{one} function, and thus not necessarily for \textit{all} functions. Typically, \textsc{esjd} is optimized for the identity function; this strategy has demonstrated on many occasions in the literature to lead to reliable conclusions (see, e.g., \cite{YANG20206094}). This choice also allows to establish a formal connection between our results and those of \cite{RobertsGelmanGilks1997} in \autoref{sec:analysis_and_tuning}.

\subsection{Notation and framework}

We first note that within our framework the Bayesian posterior $\pi$ depends on $n$; therefore, from now on the target will be denoted by $\pi_n$. The target being a posterior distribution in fact depends on a set of observations that will be denoted by $\mathbf{y}_{1:n} := (\mathbf{y}_1, \ldots, \mathbf{y}_n) \in \prod_{i=1}^n \boldsymbol{\mathsf{Y}}_i$. We make this dependence implicit to simplify. We assume $\mathbf{y}_{1:n}$ to be the first $n$ components of a realisation of some unknown data generating process $\mathbb{P}^{\mathbf{Y}}$ on $\prod_{i=1}^\infty \boldsymbol{\mathsf{Y}}_i$. Through its dependence on the data points, the distribution $\pi_n$ is a random measure on $\mathbb{R}^d$. Consequently, everything derived from it (or in fact directly from the data points) is random, such as integrals with respect to $\pi_n$ and the distributions of Markov chains produced by \gls{rwm} targeting $\pi_n$. In the following, we make statements about the convergence of such mathematical objects in $\mathbb{P}^{\mathbf{Y}}$-probability. We now briefly describe what we mean by this and refer to \citet{schmon2020a} and \citet[Section S1]{Schmon2020b} for more details on random measures and such convergences in a \gls{mcmc} context. We say, for instance, that an integral with respect to $\pi_n$, denoted by $I_n$, converges to $I$ in $\mathbb{P}^{\mathbf{Y}}$-probability when $\mathbb{P}^{\mathbf{Y}}|I_n - I| \rightarrow 0$. A Markov chain produced by \gls{rwm} targeting $\pi_n$ is seen to weakly converge in $\mathbb{P}^{\mathbf{Y}}$-probability towards another Markov chain when the finite-dimensional distributions converge in $\mathbb{P}^{\mathbf{Y}}$-probability, where the latter can be seen as random integrals involving $\pi_n$ and random transition kernels.

The matrix $\mathbf{S}$ will also depend on $n$ and will thus be written $\mathbf{S}_n$. We use $\varphi(\boldsymbol\theta; \boldsymbol\mu, \boldsymbol\Sigma)$ to denote a Gaussian density with argument $\boldsymbol\theta$, mean $\boldsymbol\mu$, and covariance matrix $\boldsymbol\Sigma$, and use $\Phi$ to denote the cumulative distribution function of a standard normal; $\F{\boldsymbol\theta}$ and $\hat{\boldsymbol\theta}_n$ denote the Fisher information evaluated at $\boldsymbol\theta$ and a parameter estimator, respectively. Finally, the norm of a vector $\boldsymbol\mu$ with respect to a matrix $\boldsymbol\Sigma$ is denoted by $\|\boldsymbol\mu\|_{\boldsymbol\Sigma}^2 := \boldsymbol\mu^T \boldsymbol\Sigma \boldsymbol\mu$. We simply write $\|\boldsymbol\mu\|^2$ when $\boldsymbol\Sigma = \mathbf{1}$.

\section{Large-sample asymptotics of RWM}
\label{sec:large_sample_asymptotic}

We first present three conditions under which a weak convergence of \gls{rwm} can be established, and next, our result.
The first condition is that a Bernstein--von Mises theorem holds, i.e.\ the concentration of the \gls{pdf} $\pi_n$ around the true model-parameter value $\boldsymbol\theta_0$, as $n$ increases, with a shape that resembles that of a Gaussian.
For simplicity, we only consider the case where the Bayesian model is well specified, but our result remains valid under model misspecification, however in this case, $\boldsymbol\theta_0$ is some fixed parameter value and the covariance matrix of the Gaussian is different \citep[see][]{kleijn2012}.

\begin{assumption}[Bernstein--von Mises theorem]
\label{assu:posterior} As $n \rightarrow \infty$, we have the following convergences in $\mathbb{P}^{\mathbf{Y}}$-probability:
\begin{align*}
&\int\left|{\pi}_{n}(\boldsymbol\theta)-\varphi(\boldsymbol\theta; \hat{\boldsymbol\theta}_{n}, \iF{\boldsymbol\theta_0}/n)\right| \mathrm{d}\boldsymbol\theta \rightarrow 0 \quad \text{with} \quad  \hat{\boldsymbol\theta}_n \rightarrow \boldsymbol\theta_0.
\end{align*}
\end{assumption}

If the posterior concentrates at a rate of $1/\surd{n}$, the scaling of the random walk needs to decrease at the same rate.
Note that this is an analogous requirement to that in \cite{RobertsGelmanGilks1997}; in that paper, the scaling diminishes with $d$ like $1/\surd{d}$. In both cases, it is to accommodate to the reality that, as $n$ or $d$  increases, the acceptance rate rapidly deteriorates if the scaling is not suitably reduced.

The scaling matrix is more precisely considered here to be of the following form: $\mathbf{S}_n = (\lambda / \surd{n}) \mathbf{M}_n$, with $\mathbf{M}_n$ a matrix that is allowed to depend on $n$ (and the data, but this dependence is made implicit to simplify the notation). The second assumption is now presented.

\begin{assumption}[Proposal scaling]
\label{assu:proposal}
The proposal is scaled as follows: $\mathbf{S}_n = (\lambda / \surd{n}) \mathbf{M}_n$, and there exists a matrix $\mathbf{M}$ such that $\mathbf{M}_n\mathbf{M}_n^T \rightarrow \mathbf{M}\mathbf{M}^T$ in $\mathbb{P}^{\mathbf{Y}}$-probability, where we say that a matrix converges in probability whenever all entries converge in probability.
\end{assumption}

A choice of matrix $\mathbf{M}_n$ that satisfies \autoref{assu:proposal} is the identity matrix $\mathbf{1}$. In the following, it will be seen that choosing $\mathbf{M}_n$ to be the result of a Cholesky decomposition of $\iF{\hat{\boldsymbol\theta}_n}$, i.e.\ such that $\mathbf{M}_n \mathbf{M}_n^T = \iF{\hat{\boldsymbol\theta}_n}$, may be preferable, depending on the strength of the correlation between the parameters.
When the correlation is significant, the desirable property is that $\mathbf{M}_n \mathbf{M}_n^T \rightarrow \mathbf{M}\mathbf{M}^T = \iF{\boldsymbol\theta_0}$ in $\mathbb{P}^\mathbf{Y}$-probability, which is often the case for regular models when $\mathbf{M}_n \mathbf{M}_n^T = \iF{\hat{\boldsymbol\theta}_n}$.
Note that other choices of matrices $\mathbf{M}_n$ may have this property. For instance, it may be valid to choose $\mathbf{M}_n$ to be the result of a Cholesky decomposition of the inverse observed information matrix instead.

Given that the target distribution concentrates and the proposal scaling decreases, we need to standardize the Markov chains simulated by \gls{rwm} to obtain a non-trivial limit. For each time step, we consider the transformation $\mathbf{z}_n := n^{1/2}(\boldsymbol\theta_n - \hat{\boldsymbol\theta}_n)$. The proposals after the transformation are thus $\mathbf{z}_n' = \mathbf{z}_n + \lambda \mathbf{M}_n \boldsymbol\epsilon$ and the resulting Markov chains have a stationary \gls{pdf} $\pi_{\mathbf{Z}_n}$ which is such that $\pi_{\mathbf{Z}_n}(\mathbf{z}_n) = \pi_n(\hat{\boldsymbol\theta}_n + n^{-1/2}\mathbf{z}_n)/n^{d/2}$.
This implies that the proposals are sampled from a Gaussian with a non-decreasing scaling and the stationary distribution behaves like a Gaussian with mean $\mathbf{0}$ and covariance $\iF{\boldsymbol\theta_0}$, as $n \rightarrow \infty$. Let $\Xi_{n} := \big(\mathbf{Z}_{k, n}\big)_{k\geqslant 0}$ be such a standardized Markov chain with $\mathbf{Z}_{k, n}$ being the state of the chain after $k$ iterations.

An asymptotic result that we prove is a convergence of $\Xi_{n}$ towards $\Xi := \big(\mathbf{Z}_{k}\big)_{k\geqslant 0}$, which is a Markov chain simulated by a \gls{rwm} algorithm targeting a Gaussian with mean $\mathbf{0}$ and covariance $\iF{\boldsymbol\theta_0}$ using proposals given by $\mathbf{z}' = \mathbf{z} + \lambda \mathbf{M} \boldsymbol\epsilon$.

To obtain the result, we assume that the chains start in stationarity. If this is not the case, the result generally still holds (at least approximatively), but for subchains formed of states with iteration indices larger than a certain threshold. Indeed, the chains produced by \gls{rwm} are irreducible and they are typically aperiodic (they are if there are positive probabilities of rejecting proposals), therefore they are typically ergodic \citep{tierney1994markov}. This implies that the chains typically reach stationarity (at least approximatively) after a large enough number of iterations.

\begin{assumption}[Stationarity]
\label{assu:stationarity}
$\Xi_{n}$ and $\Xi$ start in stationarity.
\end{assumption}
We are now ready to present the main theoretical results of this paper.

\begin{theorem}
\label{theo:mcmc_weak_limit}
Under Assumptions \ref{assu:posterior}, \ref{assu:proposal} and \ref{assu:stationarity}, we have the following convergences in $\mathbb{P}^\mathbf{Y}$-probability:
\begin{itemize}[leftmargin=1cm]
\item[(i)] $\Xi_{n}$ converges weakly to $\Xi$;

\item[(ii)] the expected acceptance probability converges,
 \begin{align*}
 &\E\left[\min\left\{1, \frac{\pi_{\mathbf{Z}_n}(\mathbf{Z}_n')}{\pi_{\mathbf{Z}_n}(\mathbf{Z}_n)}\right\}\right] \rightarrow \E\left[\min\left\{1, \frac{\varphi(\mathbf{Z}'; \mathbf{0}, \iF{\boldsymbol\theta_0})}{\varphi(\mathbf{Z}; \mathbf{0}, \iF{\boldsymbol\theta_0})}\right\}\right],
 \end{align*}
 with $\mathbf{Z}_n \sim \pi_{\mathbf{Z}_n}$, $\mathbf{Z}_n' \sim \varphi(\, \cdot \,; \mathbf{Z}_n, \lambda^2 \mathbf{M}_n \mathbf{M}_n^T)$, $\mathbf{Z} \sim \varphi(\, \cdot \,; \mathbf{0}, \iF{\boldsymbol\theta_0})$, $\mathbf{Z}' \sim \varphi(\, \cdot \,; \mathbf{Z}, \lambda^2 \mathbf{M} \mathbf{M}^T)$;

\item[(iii)] if additionally $$\mathbf{M}_n \mathbf{M}_n^T = \iF{\hat{\boldsymbol\theta}_n} \rightarrow \mathbf{M}\mathbf{M}^T = \iF{\boldsymbol\theta_0}$$ in $\mathbb{P}^\mathbf{Y}$-probability, then the \gls{esjd} converges,
$$\E\left[\|\mathbf{Z}_{k + 1, n}  - \mathbf{Z}_{k, n}\|_{\F{\hat{\boldsymbol\theta}_n}}^2 \right] \rightarrow \E\left[\|\mathbf{Z}_{k + 1} - \mathbf{Z}_k\|_{\F{\boldsymbol\theta_0}}^2 \right].$$
\end{itemize}
\end{theorem}
The proof of \autoref{theo:mcmc_weak_limit} and of all following theoretical results are deferred to \autoref{sec:proofs}. Note that, as shown in the proof, Result (iii) holds under a more general, but more technical, assumption.

\section{Tuning guidelines and analysis of the limiting RWM}\label{sec:analysis_and_tuning}

We first present in \autoref{sec:tuning} special cases of the limiting \textsc{esjd} resulting from specific choices for $\mathbf{M}$; these special cases will be seen to suggest tuning guidelines. Subsequently, we turn to an extensive analysis of the limiting \gls{rwm} in \autoref{sec:analysis} showing the relevance of these guidelines, but also the robustness of the 0.234 rule when $\mathbf{M} = \mathbf{1}$. An interesting feature of the proposed guidelines is that they are consistent with this rule. An asymptotic connection with the results of \cite{RobertsGelmanGilks1997} as $d \rightarrow \infty$ is established in \autoref{sec:connection_scaling}.

\subsection{Tuning guidelines}\label{sec:tuning}

In the same spirit as \cite{RobertsGelmanGilks1997} who optimize the speed measure of their limiting diffusion as a proxy, we propose here to optimize $$\E\left[\|\mathbf{Z}_{k + 1} - \mathbf{Z}_k\|_{\F{\boldsymbol\theta_0}}^2 \right] =: \textsc{esjd}(\lambda, \mathbf{M})$$
with respect to the tuning parameter $\lambda$, for given $\mathbf{M}$. There exists a simple expression for $\textsc{esjd}(\lambda, \mathbf{M})$ for the typical choice $\mathbf{M} = \mathbf{1}$ or when $\mathbf{M}$ results from a Cholesky decomposition of $\iF{\boldsymbol\theta_0}$, i.e.\ when $\mathbf{M} \mathbf{M}^T = \iF{\boldsymbol\theta_0}$. The expressions are provided in \autoref{cor:esjd} below, along with the expected acceptance probabilities associated with these special cases of $\mathbf{M}$.

\begin{corollary}[Formulae for \textsc{esjd} and acceptance probabilities]\label{cor:esjd}
 Assume $\Xi$ starts in stationarity and let $\boldsymbol\epsilon \sim \varphi(\,\cdot\, ; \mathbf{0}, \mathbf{1})$. If $\mathbf{M} = \mathbf{1}$,
\begin{equation}\label{eq:esjd_formula_short_sigma}
\textsc{esjd}(\lambda, \mathbf{M}) = 2 \lambda^2 \, \E\left[ \|\boldsymbol\epsilon\|_{\F{\boldsymbol\theta_0}}^2 \, \Phi\left(-\lambda \, \frac{\|\boldsymbol\epsilon\|_{\F{\boldsymbol\theta_0}}}{2}\right) \right],
\end{equation}
and the expected acceptance probability is
\begin{equation*}
	2 \, \E\left[\, \Phi\left(-\lambda \, \frac{\|\boldsymbol\epsilon\|_{\F{\boldsymbol\theta_0}}}{2}\right) \right].
\end{equation*}
If $\mathbf{M} \mathbf{M}^T = \iF{\boldsymbol\theta_0}$,
\begin{equation}\label{eq:esjd_formula_short}
	 \textsc{esjd}(\lambda, \mathbf{M}) = 2\lambda^2 \, \E\left[\|\boldsymbol\epsilon\|^2 \, \Phi\left(-\lambda \,  \frac{\|\boldsymbol\epsilon\|}{2}\right) \right]
\end{equation}
and the expected acceptance probability is
\begin{equation*}
			2 \, \E\left[\, \Phi\left(-\lambda \, \frac{\|\boldsymbol\epsilon\|}{2}\right) \right].
\end{equation*}
\end{corollary}

In general, expressions \eqref{eq:esjd_formula_short_sigma} and \eqref{eq:esjd_formula_short} in \autoref{cor:esjd} cannot be optimized analytically, but can be approximated efficiently using independent Monte Carlo sampling, and thus, numerically optimized using the resulting approximations.
We note that \eqref{eq:esjd_formula_short_sigma} and \eqref{eq:esjd_formula_short} coincide when $\F{\boldsymbol\theta_0} = \mathbf{1}$, and that in general, \eqref{eq:esjd_formula_short_sigma} depends on  $\F{\boldsymbol\theta_0}$ while \eqref{eq:esjd_formula_short} does not. This reveals that the value of $\lambda$ maximizing \eqref{eq:esjd_formula_short_sigma} is similar to that maximizing \eqref{eq:esjd_formula_short} when the model parameters are close to be \gls{iid}, but is expected to be different otherwise. %
More precisely, it is expected that the value of $\lambda$ maximizing \eqref{eq:esjd_formula_short_sigma} is small when the parameters are strongly correlated, yielding inefficient \gls{rwm} algorithms; this is confirmed in \autoref{sec:analysis}.  \autoref{cor:esjd} also reveals that, when $\mathbf{M}$ is such that $\mathbf{M} \mathbf{M}^T = \F{\boldsymbol\theta_0}^{-1}$, the optimal value for $\lambda$ is invariant to the covariance structure. %
In other words, \autoref{cor:esjd} suggests the following practical guideline: \emph{set $\mathbf{S}_n = (\lambda / \surd{n}) \mathbf{M}_n$ with $\mathbf{M}_n$ such that $\mathbf{M}_n \mathbf{M}_n^T = \iF{\hat{\boldsymbol\theta}_n}$}.
Aiming to match the proposal covariance to the target covariance has a long history in \gls{mcmc} (see, e.g., \cite{haario2001adaptive} in a context of adaptive algorithms).
To exactly match the target covariance, $\mathbf{S}_n$ is typically set to $\mathbf{S}_n = (\lambda / \surd{n}) \mathbf{1}$ and trial runs are performed to estimate the covariance. This may turn out to be ineffective when \gls{rwm} with this choice of scaling matrix performs poorly.
The guideline proposed here provides an alternative: while the matrix used to build $\mathbf{S}_n$ \textit{does not} correspond to the target covariance, it is \textit{asymptotically equivalent} to it (under the assumptions mentioned in \autoref{sec:large_sample_asymptotic}); the advantage is that this alternative can be implemented for the very first algorithm run.

In \autoref{tab:rwmh_tuning}, we present the results of a numerical optimization of $\textsc{esjd}(\lambda, \mathbf{M})$ when $\lambda = \ell / \surd{d}$ and $\mathbf{M}$ is such that $\mathbf{M} \mathbf{M}^T = \iF{\boldsymbol\theta_0}$ based on Monte Carlo samples of size 10,000,000 and a grid search, for several values of $d$. The optimization is thus with respect to $\ell$ and the optimal value is denoted by $\hat{\ell}$. Note that we have observed empirically that optimizing the \gls{ess} yields similar results.
Note also that the code to produce all numerical results is available online\footnote{See ancillary files on \url{https://arxiv.org/abs/2104.06384}.}. In \autoref{tab:rwmh_tuning}, additionally to $\hat{\ell}$, we present the acceptance rate, i.e.\ the Monte Carlo estimate of the expected acceptance probability, of the \gls{rwm} using $\hat{\ell}$. This table thus serves as guidelines to set $\ell$ in $\mathbf{S}_n = (\ell / \surd{d n}) \mathbf{M}_n$ with $\mathbf{M}_n$ such that $\mathbf{M}_n \mathbf{M}_n^T = \iF{\hat{\boldsymbol\theta}_n}$. Writing $\lambda = \ell / \surd{d}$ allows to establish a connection with the results of \cite{RobertsGelmanGilks1997} in \autoref{sec:connection_scaling}. The existence of such a connection is highlighted by the values of the optimal acceptance rates for large values of $d$.
In \autoref{sec:connection_scaling}, we establish that \gls{esjd} converges as $d \rightarrow \infty$ to the same expression which is optimized in \cite{RobertsGelmanGilks1997} and which leads within their framework to an optimal acceptance rate of $23.38\%$. From this result, we prove that the asymptotically optimal acceptance rate derived within our framework is $23.38\%$ as well.
What is remarkable is that, not only do we retrieve within our framework the same value as \cite{RobertsGelmanGilks1997} when the parameters are \gls{iid}, i.e.\ when $\iF{\boldsymbol\theta_0} = \mathbf{1}$, but the limiting optimal acceptance rate is also $23.38\%$ when $\F{\boldsymbol\theta_0} \neq \mathbf{1}$, as long as $\mathbf{M} \mathbf{M}^T = \iF{\boldsymbol\theta_0}$, which is a consequence of the invariance of \eqref{eq:esjd_formula_short}, a quality that the acceptance rate also has.

From \autoref{tab:rwmh_tuning}, we observe that when $\mathbf{M}$ is such that $\mathbf{M} \mathbf{M}^T = \iF{\boldsymbol\theta_0}$, the optimal acceptance rate is approximately 44\% for $d=1$, 35\% for $d=2$ and decreases towards 23.38\% as $d$ increases, regardless of the covariance structure. A theoretical result allows to support our numerical findings. \autoref{prop:decreasing} below states that, for fixed $\ell$, the expected acceptance probability decreases monotonically as $d$ increases, which confirms, for instance, that from $d = 1$ to $d = 2$ with $\ell = \hat{\ell} = 2.42$ fixed, the expected acceptance probability decreases.
\begin{proposition}
\label{prop:decreasing}
Let $\boldsymbol\epsilon \sim \varphi(\,\cdot\, ; \mathbf{0}, \mathbf{1})$. For $d \geq 2$,
\begin{align*}
2 \, \E\left[\, \Phi\left(-\frac{\ell}{2}\sqrt{\frac{1}{d}\sum_{i = 1}^d \epsilon_i^2}\right) \right] \leq 2 \, \E\left[\, \Phi\left(-\frac{\ell}{2}\sqrt{\frac{1}{d-1}\sum_{i = 1}^{d - 1} \epsilon_i^2}\right) \right].
\end{align*}
\end{proposition}

We finish this section by noting that for $d = 1$, the \gls{esjd} and expected acceptance probability of a \gls{rwm} targeting a Gaussian distribution have closed-form expressions \cite[see][]{sherlock2009optimal}, and can thus be optimized using these expressions.

\begin{table*}[ht]
\def~{\hphantom{0}}
\caption{Optimal value for $\ell$ and the acceptance rate of the limiting \gls{rwm} using this value and $\mathbf{M}$ such that $\mathbf{M} \mathbf{M}^T = \iF{\boldsymbol\theta_0}$, as a function of $d$}
 \small
 \centering
{
\begin{tabular}{l rrrrrrrrrr}
\toprule
$d$                & 1     & 2    & 3    & 4     & 5     & 10    & 15    & 20    & 30    & 50 \cr
\midrule
$\hat{\ell}$       & 2.42  & 2.42 & 2.42 &  2.42 & 2.40  & 2.40  & 2.39  & 2.39  & 2.38  & 2.38 \cr
Acc. rate. (in \%) & 44.00 & 35.00 & 31.30 & 29.29 & 28.39 & 25.78 & 25.07 & 24.61 & 24.34 & 23.97 \cr
 \bottomrule
\end{tabular}
}
\label{tab:rwmh_tuning}
\end{table*}

\subsection{Analysis of the limiting RWM}\label{sec:analysis}

We now present the practical implications of the guidelines proposed in \autoref{sec:tuning} (in the asymptotic regime $n \rightarrow \infty$) through an analysis of the impact of different target covariances on the performance and acceptance rate of the optimal limiting \gls{rwm}. More precisely, we analyse the behaviour of the limiting \gls{rwm} with $\mathbf{M} = \mathbf{1}$ and $\mathbf{M}$ such that $\mathbf{M} \mathbf{M}^T = \iF{\boldsymbol\theta_0}$ under different target covariances; for each of these covariances, the algorithms are made optimal, in the sense that $\lambda$ (or $\ell$) is tuned according to the expressions in \autoref{cor:esjd} (or \autoref{tab:rwmh_tuning}). The algorithm with $\mathbf{M}$ such that $\mathbf{M} \mathbf{M}^T = \iF{\boldsymbol\theta_0}$ has a higher complexity because an additional matrix multiplication is required every iteration. However, in standard modern statistical-computing frameworks we found both algorithms to take roughly the same amount of time to complete; it is the case for instance for the numerical experiments presented in this paper that were performed in \textsf{R} \citep{rprogram} on a computer with an i9 CPU.

\begin{figure*}[ht]
  \centering
  $\begin{array}{cc}
 \hspace{-2mm}\includegraphics[width=0.50\textwidth]{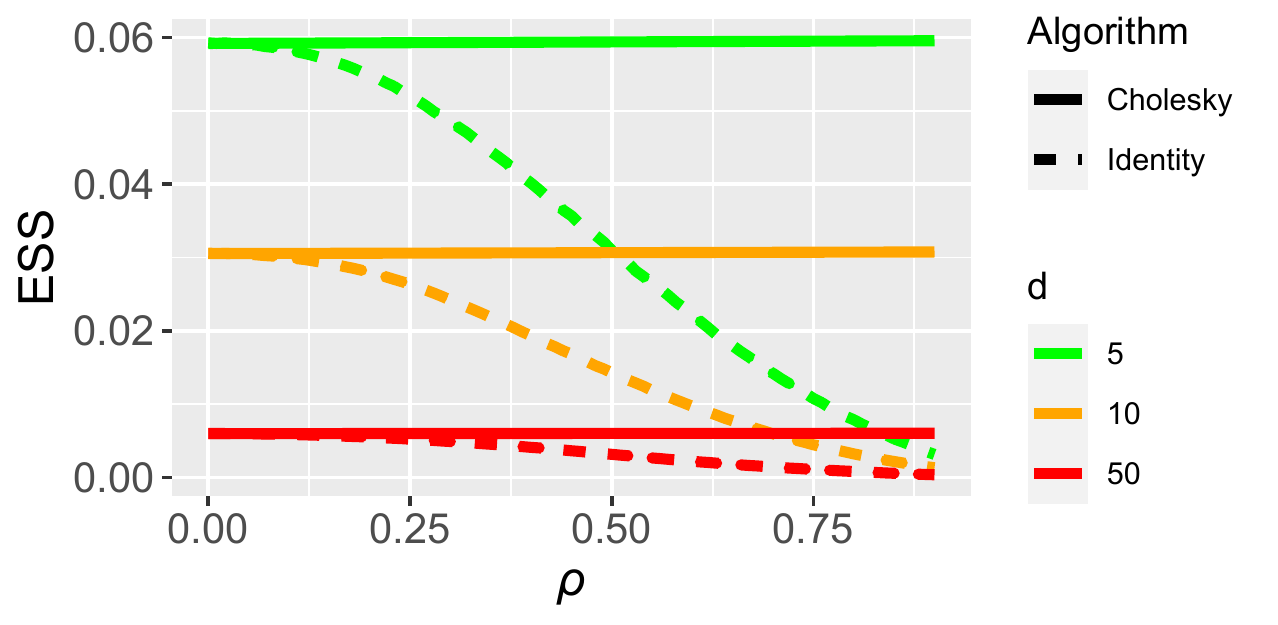} &  \hspace{-3mm} \includegraphics[width=0.50\textwidth]{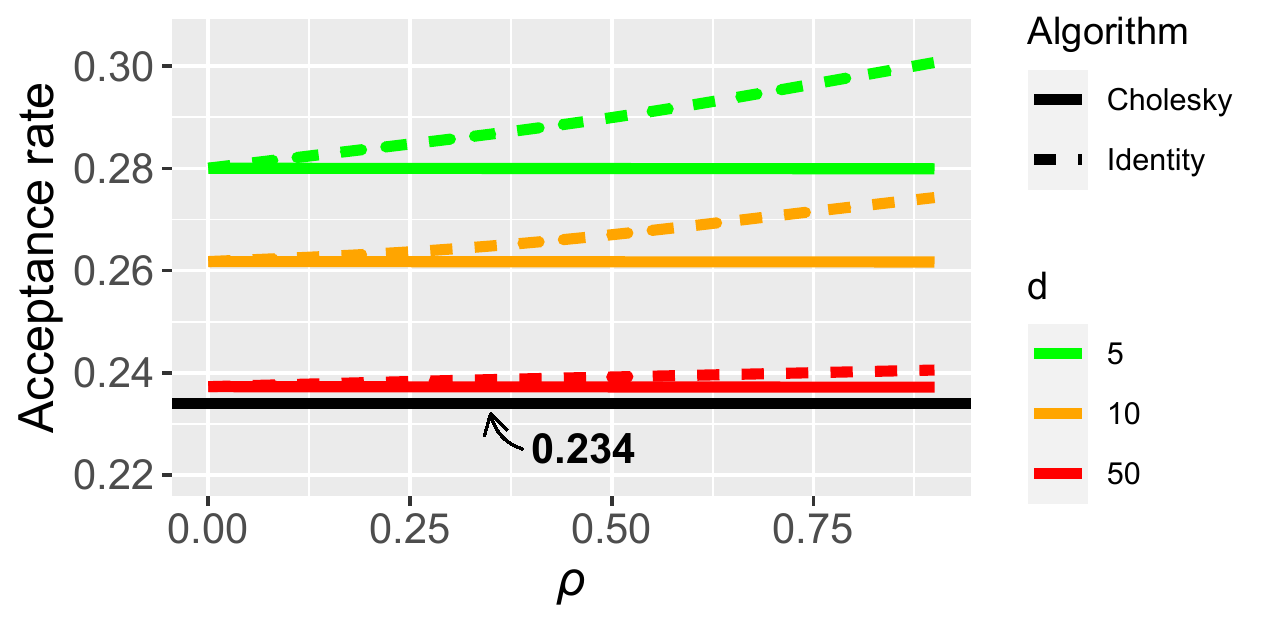} \cr
   \hspace{-14mm} \textbf{(a) } & \hspace{-14mm} \textbf{(b) } \cr
  \end{array}$\vspace{-2mm}
  \caption{\small Optimal (a) \gls{ess} and (b) acceptance rate of the limiting \gls{rwm} with $\mathbf{M} = \mathbf{1}$ and with $\mathbf{M}$ such that $\mathbf{M} \mathbf{M}^T = \iF{\boldsymbol\theta_0}$ as a function of $\rho$ in the case where the $(i,j)$th entry of $\iF{\boldsymbol\theta_0}$ is given by $\rho^{|i-j|}$, when $d = 5, 10, 50$}\label{fig:impact_rho}
 \end{figure*}

For the analysis, we focus on showing what happens when the correlation between the model parameters increases under a specific covariance structure: the $(i,j)$th entry of $\iF{\boldsymbol\theta_0}$ is given by $\rho^{|i-j|}$, where $-1 \leq \rho \leq 1$ is a varying parameter.
This covariance structure is often called \textit{autoregressive of order 1} and represents a situation where the parameters are standardized, in the sense that their marginal variances are all equal to 1, and the correlations between them decline exponentially with distance, at a speed that depends on $\rho$.
In this setting, the target covariance matrix is parametrized with only one parameter, $\rho$. The case where $0 \leq \rho \leq 1$ is more interesting for the current purpose; a value close to 0 leads to weak correlations between the parameters, whereas a value close to 1 makes the correlation persist with distance, yielding strong correlations between the parameters. Note that the situation where parameters are standardized and $\mathbf{M} = \mathbf{1}$ is equivalent to that where the parameters are non-standardized but $\mathbf{M}$ is a diagonal matrix with diagonal entries equal to the marginal standard deviations. The empirical results are presented in \autoref{fig:impact_rho}.

In \autoref{fig:impact_rho}, the algorithm performances are evaluated using the minimum of the marginal \gls{ess}s, reported per iteration. \gls{esjd} cannot be used to evaluate performance across different values of $\rho$ because using a norm with respect to $\F{\boldsymbol\theta_0}$ in \gls{esjd} standardizes this measure. We show the results for $0 \leq \rho \leq 0.9$ as beyond $0.9$, \gls{rwm} with $\mathbf{M} = \mathbf{1}$ becomes unreliable. As suggested by the expressions in \autoref{cor:esjd}, the performance of \gls{rwm} with $\mathbf{M}$ such that $\mathbf{M} \mathbf{M}^T = \iF{\boldsymbol\theta_0}$ does not vary with $\rho$, while it does for \gls{rwm} with $\mathbf{M} = \mathbf{1}$; it in fact deteriorates when $\rho$ increases due to an optimal value for $\ell$ that decreases. As for the acceptance rate, it is invariant as well for \gls{rwm} with the Cholesky-decomposition matrix, and increases slightly with $\rho$ for \gls{rwm} with the identity matrix. The optimal acceptance rate becomes closer to $0.234$ as $d$ increases when $\rho = 0$, which is not surprising given that the target in this case satisfies the assumptions of \cite{RobertsGelmanGilks1997}. It is however remarkable that, for $\mathbf{M} = \mathbf{1}$, the optimal acceptance rate only slightly increases as $\rho$ gets closer to 1.

\normalsize

\subsection{Connection to scaling limits}\label{sec:connection_scaling}

The aim of this section is to establish a formal connection between our guidelines and those of \citet{RobertsGelmanGilks1997}
through an asymptotic analysis of features of the limiting chain $\Xi := \big(\mathbf{Z}_{k}\big)_{k\geqslant 0}$ as $d$ increases.
In particular, it will be pointed out using a theoretical argument that our guidelines are consistent in that we find equivalent asymptotically optimal values for $\ell$ and acceptance rate as these authors. The stationary distribution of $\Xi$, which is a Gaussian with mean $\mathbf{0}$ and covariance $\iF{\boldsymbol\theta_0}$, can be seen as a special case of the product target studied by \citet{RobertsGelmanGilks1997} when $\iF{\boldsymbol\theta_0} = \mathbf{1}$. As mentioned in the previous sections, it is thus not surprising but reassuring to find the same asymptotically optimal values within our framework for this special case.

To find the optimal values for \gls{rwm} in the high-dimensional limit, we analyse the expected acceptance probability and $\textsc{esjd}(\lambda, \mathbf{M})$ by considering them as sequences indexed by $d$, and let $d \rightarrow \infty$. We provide a result establishing that $\textsc{esjd}(\lambda, \mathbf{M})$ converges towards a function that is equivalent to that optimized in \citet{RobertsGelmanGilks1997}, when $\lambda = \ell/\surd{d}$ and the proposal covariance is set to $\mathbf{M} \mathbf{M}^T = \iF{\boldsymbol\theta_0}$. The \gls{esjd} is optimized by an equivalent value for $\ell$, and the expected acceptance probability converges to the same limiting acceptance rate as \citet{RobertsGelmanGilks1997}, which is seen to imply that the asymptotically optimal acceptance rate is the same. The asymptotically optimal values are $2.38$ and $0.234$ for $\ell$ and the acceptance rate, respectively. Within our framework, these values are optimal for any target covariance $\iF{\boldsymbol\theta_0}$ given that the limiting acceptance rate and \textsc{esjd} do not depend on $\iF{\boldsymbol\theta_0}$.

Before presenting the formal results, we provide an informal argument explaining why the connection exists and more precisely why $\textsc{esjd}(\lambda, \mathbf{M})$ converges towards a function that is equivalent to that in \citet{RobertsGelmanGilks1997}. Central to the reason why the efficiency measures are asymptotically the same are the convergences of the acceptance rates in both contexts to a constant as $d \rightarrow \infty$. To provide the informal argument, we thus present the acceptance rates and show how Taylor expansions explain their asymptotic behaviour. We start with that in \citet{RobertsGelmanGilks1997}; we thus consider a sequence of target densities $\{\pi_d\}$ with $\pi_d(\btheta) = \prod_{i=1}^d f(\theta_i)$ and $\btheta' = \btheta + (\ell / \surd{d}) \boldsymbol\epsilon$, $f$ satisfying some regularity conditions. Under these assumptions, it can be proved that for $d$ large,
\begin{align}
 \E\left[\min\left\{1, \frac{\pi_d(\btheta')}{\pi_d(\btheta)}\right\}\right] &\approx \E\left[\min\left\{1, \exp\left(\sum_{i=1}^d \psi(\theta_i)(\theta_i' - \theta_i) - \frac{\ell^2}{2d}\psi(\theta_i)^2\right)\right\}\right] \nonumber \\
  &= 2\E\left[\Phi\left(-\frac{\ell}{2}\sqrt{\frac{1}{d}\sum_{i=1}^d \psi(\theta_i)^2}\right)\right], \label{eqn:arg_3_3}
\end{align}
where ``$\approx$'' is to be understood as a relationship asserting that the expressions are asymptotically equivalent and
\[
 \psi(\theta_i) := \left. \frac{\partial}{\partial x} \log f(x)\right|_{x=\theta_i};
\]
for the equality \eqref{eqn:arg_3_3}, we used that the term in the exponential has a conditional normal distribution given $\btheta$ (because $\theta_i' - \theta_i = (\ell / \surd{d}) \epsilon_i$) and the closed-form of $\E[\min\{1, e^X\}]$ when $X \sim \varphi$.
We establish a limit using that
\[
 2\E\left[\Phi\left(-\frac{\ell}{2}\sqrt{\frac{1}{d}\sum_{i=1}^d \psi(\theta_i)^2}\right)\right] \rightarrow 2\Phi(-\ell \surd{L}/2),
\]
with $$L := \E[\psi(\theta_1)^2].$$  In their context, $\hat{\ell} = 2.38 / \surd{L}$ and $2 \, \Phi\left(-\hat{\ell}\surd{L}/2\right) = 0.234$.

In our framework, we first consider a sequence of posterior densities $\{\pi_n\}$ based on observations of \gls{iid} random variables $\mathbf{Y}_i \sim g_{\btheta}$, $g_{\btheta}$ satisfying some regularity conditions. Under Assumptions \ref{assu:posterior} and \ref{assu:proposal} and setting $\mathbf{S}_n = (\ell / \surd{d n}) \mathbf{M}_n$ with $\mathbf{M}_n\mathbf{M}_n^T = \iF{\hat{\boldsymbol\theta}_n}$, it can be proved that for $n$ large:
\begin{align*}
 \E\left[\min\left\{1, \frac{\pi_n(\btheta')}{\pi_n(\btheta)}\right\}\right] &= \E\left[\min\left\{1, \frac{\pi_n(\hat{\boldsymbol\theta}_n + n^{-1/2}\mathbf{Z}'_n)}{\pi_n(\hat{\boldsymbol\theta}_n + n^{-1/2}\mathbf{Z}_n)}\right\}\right] \cr
 &\approx \E\left[\min\left\{1, \frac{\pi_n(\boldsymbol\theta_0 + n^{-1/2}\mathbf{Z}'_n)}{\pi_n(\boldsymbol\theta_0 + n^{-1/2}\mathbf{Z}_n)}\right\}\right] \cr
 &\approx \E\left[\min\left\{1, \exp\left(-\frac{1}{2}\|\mathbf{Z}'_n\|_{\hat{\mathcal{I}}_n(\btheta_0)}^2 + \frac{1}{2}\|\mathbf{Z}_n\|_{\hat{\mathcal{I}}_n(\btheta_0)}^2\right)\right\}\right],
\end{align*}
where
\[
 \hat{\mathcal{I}}_n(\btheta_0) := \frac{1}{n}\sum_{i=1}^n-\left.\frac{\partial^2}{\partial\btheta \partial\btheta^T} \log g_{\btheta}(\mathbf{y}_i)\right|_{\btheta = \btheta_0},
\]
using that $\hat{\boldsymbol\theta}_n \rightarrow \boldsymbol\theta_0$ and that the local asymptotic normality allows an expansion of $\log \pi_n(\boldsymbol\theta_0 + n^{-1/2}\mathbf{z}_n)$ with vanishing terms beyond order 2. The last expectation above is asymptotically equivalent to
\[
 \E\left[\min\left\{1, \frac{\varphi(\mathbf{Z}'; \mathbf{0}, \iF{\boldsymbol\theta_0})}{\varphi(\mathbf{Z}; \mathbf{0}, \iF{\boldsymbol\theta_0})}\right\}\right],
\]
with $\mathbf{Z} \sim \varphi(\, \cdot \,; \mathbf{0}, \iF{\boldsymbol\theta_0})$ and $\mathbf{Z}' \sim \varphi(\, \cdot \,; \mathbf{Z}, \lambda^2 \mathbf{M} \mathbf{M}^T)$. The latter expectation is equal to (recall \autoref{cor:esjd})
\[
 2 \, \E\left[ \Phi\left(- \frac{\ell \|\boldsymbol\epsilon\|}{2 \surd{d}}\right) \right]  \rightarrow 2 \, \Phi\left(-\frac{\ell}{2}\right),
\]
 as $d \rightarrow \infty$. When $\mathbf{M}_n\mathbf{M}_n^T = \iF{\hat{\boldsymbol\theta}_n}$, $L = 1$ because the proposal covariance is set to asymptotically match the target covariance exactly and thus $\hat{\ell} = 2.38$ with $2 \, \Phi\left(-\hat{\ell}/2\right) = 0.234$.
If, alternatively, the proposal is set to an isotropic Gaussian, i.e. $\mathbf{M}_n = \mathbf{1}$, a constant analogous to $L$ appears in the limiting acceptance rate:
$$
	L' := \lim_{d\rightarrow \infty} \frac{\|\boldsymbol\epsilon\|_{\F{\boldsymbol\theta_0}}^2}{d},
$$
provided that this limit exists (in distribution).

The formal results are presented in \autoref{prop:scaling_const}.

\begin{proposition}[Guideline consistency] \label{prop:scaling_const}
If $\Xi$ starts in stationarity, $\lambda = \ell/\surd{d}$ and $\mathbf{M} \mathbf{M}^T = \iF{\boldsymbol\theta_0}$, then
\begin{align*}
	\textsc{esjd}(\lambda, \mathbf{M}) &:= \E\left[\|\mathbf{Z}_{k + 1} - \mathbf{Z}_k\|_{\F{\boldsymbol\theta_0}}^2 \right] = 2\ell^2 \, \E\left[ \frac{\|\boldsymbol\epsilon\|^2}{d} \, \Phi\left(- \frac{\ell \|\boldsymbol\epsilon\|}{2 \surd{d}}\right) \right] \rightarrow 2\ell^2 \, \Phi\left(- \frac{\ell}{2}\right),
\end{align*}
and
\begin{align*}
  &\E\left[\min\left\{1, \frac{\varphi(\mathbf{Z}'; \mathbf{0}, \iF{\boldsymbol\theta_0})}{\varphi(\mathbf{Z}; \mathbf{0}, \iF{\boldsymbol\theta_0})}\right\}\right] = 2 \, \E\left[ \Phi\left(- \frac{\ell \|\boldsymbol\epsilon\|}{2 \surd{d}}\right) \right]  \rightarrow 2 \, \Phi\left(- \frac{\ell}{2}\right),
\end{align*}
as $d \rightarrow \infty$, with $\mathbf{Z} \sim \varphi(\, \cdot \,; \mathbf{0}, \iF{\boldsymbol\theta_0})$ and $\mathbf{Z}' \sim \varphi(\, \cdot \,; \mathbf{Z}, \lambda^2 \mathbf{M} \mathbf{M}^T)$. Viewed as a function of $\ell$, $2\ell^2 \, \Phi\left(- \ell / 2\right)$ is maximized by $\ell = \hat{\ell} := 2.38$, and we obtain $2 \, \Phi\left(-\hat{\ell}/2\right) = 0.234$.
\end{proposition}

In theory, one can obtain a more general limiting expression for $\textsc{esjd}(\lambda, \mathbf{M})$ when $\mathbf{M}$ is not specified to be such that $\mathbf{M} \mathbf{M}^T = \iF{\boldsymbol\theta_0}$. However, one would need to know how $\iF{\boldsymbol\theta_0}$ behaves when $d$ grows because $\textsc{esjd}(\lambda, \mathbf{M})$ depends, in general, on $\iF{\boldsymbol\theta_0}$. 
For example, from \eqref{eq:esjd_formula_short_sigma}, it can be observed that
\begin{equation*}
2 \ell^2 \, \E\left[\frac{\|\boldsymbol\epsilon\|_{\F{\boldsymbol\theta_0}}^2}{d} \, \Phi\left(-\frac{\ell \|\boldsymbol\epsilon\|_{\F{\boldsymbol\theta_0}}}{2 \surd{d}}\right) \right] \rightarrow 2\ell^2 L' \, \Phi\left(-\frac{\ell\surd L'}{2} \right),
\end{equation*}
whenever $\|\boldsymbol\epsilon\|_{\F{\boldsymbol\theta_0}}^2 / d \rightarrow L'\in\mathbb{R}$ as $d \rightarrow \infty$ in probability, that is, whenever the correlation in $\F{\boldsymbol\theta_0}$ allows for a law of large numbers of the squared norm $\|\boldsymbol\epsilon\|_{\F{\boldsymbol\theta_0}}^2$, as long as uniform integrability conditions hold.
In the previous section, for example, the autoregressive covariance matrix allows for a law of large numbers and uniform integrability conditions hold. This is a consequence of the form of $\F{\boldsymbol\theta_0}$, which is a tridiagonal matrix, turning $\|\boldsymbol\epsilon\|_{\F{\boldsymbol\theta_0}}^2$ into a sum of correlated random variables, but where the correlation exists only for random variables that are close to each other; more precisely, each random variable in the sum is correlated with those with indices that differ by $1$. The conditions aforementioned may fail to hold when the matrix $\F{\boldsymbol\theta_0}$ yields a sum of correlated random variables where each of them is correlated to a number of random variables that grows with $d$.

The limiting behaviour of \gls{esjd} for the case $\mathbf{M} = \mathbf{1}$ recently received detailed attention in \citet{YANG20206094}. These authors perform analyses under the traditional asymptotic framework $d\rightarrow \infty$; however, in contrast to earlier work, their approach does not require the restrictive assumption of \gls{iid} model parameters. Instead, the authors perform analyses under an assumption of partially connected graphical models. A key mathematical object there which measures the ``roughness'' of the log target density is
\begin{equation*}
	I_d(\theta) := \frac{1}{d} \sum_{i=1}^d \left(\frac{\partial}{\partial \theta_i} \log \pi_d(\btheta)\right)^2.
\end{equation*}
It appears, for instance, in an expectation that is asymptotically equivalent to their expected acceptance probability:
\begin{equation}\label{eqn:acc_Yang}
 2 \E\left[\Phi\left(- \frac{\ell}{2} \sqrt{I_d(\btheta)}\right)\right],
\end{equation}
where the expectation is with respect to $\pi_d$. It also appears in an expectation analogous to \eqref{eq:esjd_formula_short_sigma} that is asymptotically equivalent to their \gls{esjd}. There exists an interesting connection between their optimization problem and that of optimizing \eqref{eq:esjd_formula_short_sigma} that can be established by identifying the counterpart to $I_d(\theta)$ in \eqref{eq:esjd_formula_short_sigma} and the expected acceptance probability. The optimal acceptance rates derived under their framework are often close to $0.234$, for large enough $d$, which is what we observed under our framework as well, for instance, in \autoref{sec:analysis}. We finish this section with a brief analysis which highlights the existence of that connection by focussing on similarities in between the acceptance rates.

We identify the counterpart to $I_d(\btheta)$ to be
\begin{align*}
\frac{\|\boldsymbol\epsilon\|_{\F{\boldsymbol\theta_0}}^2}{d}
&= \frac{1}{d}\sum_{i=1}^d\sum_{j=1}^d \epsilon_i\epsilon_j \F{\boldsymbol\theta_0}_{ij},
\end{align*}
recalling that
\[
 \F{\boldsymbol\theta}_{ij} = \E\left[\left(\frac{\partial}{\partial \theta_i}\log g_{\btheta}(\mathbf{Y})\right)\left(\frac{\partial}{\partial \theta_j}\log g_{\btheta}(\mathbf{Y})\right)\right].
\]
Note that under regularity conditions, the normalized version of $\left(\frac{\partial}{\partial \theta_i} \log \pi_d(\btheta)\right)^2$, when seen as the square of the derivative of the sum of the log prior and log densities, converges in distribution to $\F{\boldsymbol\theta}_{ii}$ times a chi-square random variable with $1$ degree of freedom as $n \rightarrow \infty$. For weak interactions in between model parameters represented by sparse graphs, $\|\boldsymbol\epsilon\|_{\F{\boldsymbol\theta_0}}^2 / d$ thus encodes similar information to $I_d(\theta)$. This highlights that the expected acceptance probability under our framework, given by
\[
 2 \, \E\left[ \Phi\left(- \frac{\ell \|\boldsymbol\epsilon\|_{\F{\theta_0}}}{2 \surd{d}}\right) \right],
\]
and theirs, given by \eqref{eqn:acc_Yang}, are similar in essence. In general, Jensen's inequality allows to observe that
\begin{equation*}
2 \, \E\left[ \Phi\left(- \frac{\ell \|\boldsymbol\epsilon\|_{\F{\theta_0}}}{2 \surd{d}}\right) \right] \geq 2 \,  \Phi\left(- \frac{\ell}{2} \sqrt{\frac{1}{d}\sum_{i=1}^d \F{\boldsymbol\theta_0}_{ii}}\right),
\end{equation*}
given that $x \mapsto \Phi(- a \sqrt{x})$ is convex for $x \geq 0$ with $a > 0$. Acceptance rates derived within our framework are thus expected to be larger than those derived within the framework of \citet{YANG20206094}, when $\pi_d$ concentrates around $\btheta_0$. They have for instance been observed to be larger than $0.234$ in \autoref{sec:analysis},  while in \citet{YANG20206094} they are shown to be smaller than or equal to $0.234$.

We do not investigate the problem of convergence of $\textsc{esjd}(\lambda, \mathbf{M})$ in full generality. Additionally to \citet{YANG20206094}, we refer the reader to \cite{ghosal2000asymptotic}, \cite{belloni2009computational} and \cite{belloni2014posterior} who conducted analyses of posterior distributions in asymptotic regimes where $d$ is allowed to grow with $n$.

\section{Logistic regression with real data}
\label{sec:logistic_regression}

In this section, we demonstrate that the \gls{rwm} algorithm targeting $\pi_n$ behaves similarly to its asymptotic counterpart, targeting a Gaussian distribution, in some practical cases. To achieve this, we consider a specific practical case and compare the asymptotically optimal value for $\ell$ when $\mathbf{M} \mathbf{M}^T = \iF{\boldsymbol\theta_0}$ based on \gls{esjd} (which does not depend on the unknown $\iF{\boldsymbol\theta_0}$) to that obtained from tuning the non-limiting \gls{esjd} with $\mathbf{M}_n \mathbf{M}_n^T$ set to be the inverse of the observed information matrix. We also compare the optimal acceptance rates and present results for the \gls{rwm} algorithm using $\mathbf{M}_n = \mathbf{1}$. The practical case that we study is one where the posterior distribution results from a Bayesian logistic regression model and a patent data set from \citet{fahrmeir2007regression}. We will see that for this example with a sample size of $n = 4{,}866$ and $d = 9$ parameters, both the optimal values for $\ell$ and acceptance rates coincide accurately, showing that the limiting \gls{rwm} represents a good approximation of that targeting $\pi_n$ in situations where the Bayesian models are regular and the sample sizes are realistically large. This example also allows to show that the guidelines derived from the limiting \gls{rwm} and the performance analysis conducted in \autoref{sec:analysis} are relevant in such situations.

We denote the binary-response-variable and covariate-vector data points by $r_1, \ldots, r_n$ and $\mathbf{x}_1, \ldots, \mathbf{x}_n$, respectively, with the first component of each $\mathbf{x}_i$ being equal to 1. In logistic regression, the parameters $\boldsymbol\theta$ are regression coefficients. Let us assume that $\mathbf{Y}_1, \ldots, \mathbf{Y}_n  = (R_1, \mathbf{X}_1), \ldots, (R_n, \mathbf{X}_n)$ are \gls{iid} random variables and also that the model is well specified in order to fit in the theoretical framework presented in \autoref{sec:large_sample_asymptotic}. Formally speaking, the latter assumption is certainly not true, but the fact that the empirical results are close to the theoretical (and asymptotic) ones suggests that the model approximates well the true data-generating process. We now show that \autoref{theo:mcmc_weak_limit} can be applied by verifying the assumptions stated in \autoref{sec:large_sample_asymptotic}. The logistic regression model is, as mentioned in \autoref{sec:contributions},  regular enough; \autoref{assu:posterior} is thus satisfied. We set $\mathbf{M}_n \mathbf{M}_n^T$ to be the inverse of a standardized version of the observed information matrix evaluated at the maximum a posteriori estimate $\hat{\boldsymbol\theta}_n$, i.e.\ the inverse of
\begin{align}\label{eqn:obs_inf_example}
 \frac{1}{n} \sum_{i = 1}^n \mathbf{x}_i \mathbf{x}_i^T p_i(\hat{\boldsymbol\theta}_n)(1 - p_i(\hat{\boldsymbol\theta}_n)), \intertext{where} p_i(\hat{\boldsymbol\theta}_n) := \frac{\exp(\mathbf{x}_i^T \hat{\boldsymbol\theta}_n)}{1 + \exp(\mathbf{x}_i^T \hat{\boldsymbol\theta}_n)}. \nonumber
\end{align}
Under weak regularity conditions, $\mathbf{M}_n \mathbf{M}_n^T$ converges and we set $\mathbf{S}_n = (\lambda / \surd{n}) \mathbf{M}_n$, implying that \autoref{assu:proposal} is satisfied if these weak regularity conditions are verified. \autoref{theo:mcmc_weak_limit} therefore holds provided that the chains start in stationarity (\autoref{assu:stationarity}) and these weak regularity conditions are verified.

When $d = 9$, the asymptotically optimal value for $\ell$ when $\mathbf{M} \mathbf{M}^T = \iF{\boldsymbol\theta_0}$ is $2.39$ and the acceptance rate of the limiting \gls{rwm} using this value is $26.26\%$. The optimal values for the \gls{rwm} algorithm with $\mathbf{M}_n$ set as the inverse of \eqref{eqn:obs_inf_example} are essentially the same: $2.37$ and $26.68\%$ for $\ell$ and the acceptance rate, respectively. The value of $\ell$ that maximizes the \gls{ess} per iteration is $2.40$; the maximum \gls{ess} per iteration is $0.034$, which is significantly higher than the maximum of $0.006$ attained by the algorithm with $\mathbf{M}_n = \mathbf{1}$. As explained and shown in \autoref{sec:analysis_and_tuning}, a poor performance of the latter sampler is due to strong correlation in between the parameters. For this sampler, a value of $6.89$ is optimal for $\ell$ based on the \gls{ess}, whereas a value of $6.51$ is optimal when the \gls{esjd} is instead considered. The acceptance rate of the algorithm using $\mathbf{M}_n = \mathbf{1}$ and the latter value is $27.69\%$. Note that we tried smaller models with less covariates and larger ones with  interaction terms, and the optimal values when $\mathbf{M}_n$ is set as the inverse of \eqref{eqn:obs_inf_example} are consistent with the guidelines presented in \autoref{tab:rwmh_tuning}. The results in this numerical experiment follow from a numerical optimization of \gls{esjd} and \gls{ess} based on Markov-chain samples of size 10,000,000 and a grid search.

\section{Discussion}

\glsresetall

In this paper, we have analysed the behaviour of \gls{rwm} algorithms when used to sample from Bayesian posterior distributions, under the asymptotic regime $n \rightarrow \infty$, in contrast with previous asymptotic analyses where $d\rightarrow \infty$. Our analysis led to novel parameter-dimension-dependent tuning guidelines which are consistent with the well-known $0.234$ rule. A formal argument allowed to show that this rule can in fact be derived from the angle adopted in this paper as well. We believe that similar analyses to those performed in this paper can be conducted to develop practical tuning guidelines for more sophisticated algorithms like Metropolis-adjusted Langevin algorithm \citep{roberts1996exponential} and Hamiltonian Monte Carlo \citep{Duane1987}, and to establish other interesting connections with optimal-scaling literature \citep[e.g.][]{Roberts1998, beskos2013optimal}.

The guidelines developed in this paper for \gls{rwm} algorithms are valid under weak assumptions; we essentially only require a Bernstein--von Mises theorem to hold for the target distribution. This is in stark contrast to scaling-limit approaches. To our knowledge, there is one contribution, \cite{YANG20206094}, that provides guidelines for a realistic model based on a scaling-limit argument, and it requires the posterior distribution to concentrate, which is in line with the argument of this paper. The guidelines proposed in our paper are in theory valid in the limit $n \rightarrow \infty$; we have demonstrated that they are nevertheless applicable in realistic scenarios with typical data sizes using an example of logistic-regression analysis of real data. This example, together with our analysis of the limiting \gls{rwm}, also allow to support the findings about the robustness of the $0.234$ rule to non-\gls{iid} model parameters when the scaling matrix is a diagonal matrix.

\appendix

\allowdisplaybreaks

\section{Proofs}\label{sec:proofs}

\begin{proof}[Proof of \autoref{theo:mcmc_weak_limit}]

\emph{Result (i).} To prove this result, we use Theorem 2 of  \cite{Schmon2020}. We thus have to verify three conditions.

\begin{enumerate}[leftmargin=1cm]
 \item As $n \rightarrow \infty$, the following convergence holds in $\mathbb{P}^\mathbf{Y}$-probability: $\mathbf{Z}_{0, n}$ converges weakly to $\mathbf{Z}_{0}$.

 \item Use $P_n$ and $P$ to denote the transition kernels of $\Xi_{n}$ and $\Xi$, respectively. These are such that
 \[
  \int |P_nh(\mathbf{z}) - Ph(\mathbf{z})| \, \pi_{\mathbf{Z}_n}(\mathbf{z}) \, \mathrm{d}\mathbf{z} \rightarrow 0,
 \]
 in $\mathbb{P}^\mathbf{Y}$-probability as $n \rightarrow \infty$, for all $h \in \text{BL}$, the set of bounded Lipschitz functions.
 \item The transition kernel $P$ is such that $Ph(\, \cdot \,)$ is continuous for any $h \in \mathcal{C}_{\text{b}}$, the set of continuous bounded functions.
\end{enumerate}

We start with Condition 1. It suffices to verify that
\[
 |\Prob(\mathbf{Z}_{0, n} \in A) - \Prob(\mathbf{Z}_{0} \in A)| \rightarrow 0,
\]
in $\mathbb{P}^\mathbf{Y}$-probability, for any measurable set $A$. We have that
\[
 |\Prob(\mathbf{Z}_{0, n} \in A) - \Prob(\mathbf{Z}_{0} \in A)| \leq \int\left|{\pi}_{n}(\boldsymbol\theta)-\varphi(\boldsymbol\theta; \hat{\boldsymbol\theta}_{n}, \iF{\boldsymbol\theta_0}/n)\right| \mathrm{d}\boldsymbol\theta \rightarrow 0
\]
in $\mathbb{P}^\mathbf{Y}$-probability by \autoref{assu:posterior}, using Jensen's inequality, that $A \subseteq \R^d$, and a change of variable $\boldsymbol\theta = \mathbf{z} / n^{1/2} +  \hat{\boldsymbol\theta}_n$.

We turn to Condition 2. We have that
\[
 P_n(\bz, \d\bz') = \alpha_n(\bz, \bz') \, \varphi(\d\bz'; \bz, \lambda^2 \mathbf{M}_n \mathbf{M}_n^T) + \rho_n(\bz) \, \delta_{\bz}(\d\bz'),
\]
and
\[
 P(\bz, \d\bz') = \alpha(\bz, \bz') \, \varphi(\d\bz'; \bz, \lambda^2 \mathbf{M} \mathbf{M}^T) + \rho(\bz) \, \delta_{\bz}(\d\bz'),
\]
where here
\[
 \alpha_n(\bz, \bz') := \min\left\{1, \frac{\pi_{\mathbf{Z}_n}(\bz')}{\pi_{\mathbf{Z}_n}(\bz)}\right\},
\]
$\rho_n(\bz)$ is the corresponding rejection probability, and
\[
 \alpha(\bz, \bz') := \min\left\{1, \frac{\varphi(\bz'; \mathbf{0}, \iF{\boldsymbol\theta_0})}{\varphi(\bz; \mathbf{0}, \iF{\boldsymbol\theta_0})}\right\},
\]
$\rho(\bz)$ is the corresponding rejection probability. Thus,
\[
 P_nh(\mathbf{z}) = \int h(\bz') \, \alpha_n(\bz, \bz') \, \varphi(\d\bz'; \bz, \lambda^2 \mathbf{M}_n \mathbf{M}_n^T) + h(\bz) \, \rho_n(\bz),
\]
and
\[
 Ph(\mathbf{z}) = \int h(\bz') \, \alpha(\bz, \bz') \, \varphi(\d\bz'; \bz, \lambda^2 \mathbf{M} \mathbf{M}^T) + h(\bz) \, \rho(\bz).
\]
Therefore, using the triangle inequality,
\begin{align*}
 &\int |P_nh(\mathbf{z}) - Ph(\mathbf{z})| \, \pi_{\mathbf{Z}_n}(\mathbf{z}) \, \mathrm{d}\mathbf{z} \cr
  &\quad \leq \int \left|\int h(\bz') \, \alpha_n(\bz, \bz') \, \varphi(\d\bz'; \bz, \lambda^2 \mathbf{M}_n \mathbf{M}_n^T) - \int h(\bz') \, \alpha(\bz, \bz') \, \varphi(\d\bz'; \bz, \lambda^2 \mathbf{M} \mathbf{M}^T)\right| \, \pi_{\mathbf{Z}_n}(\mathbf{z}) \, \mathrm{d}\mathbf{z} \cr
  &\qquad + \int |h(\bz) \, \rho_n(\bz) - h(\bz) \, \rho(\bz)| \, \pi_{\mathbf{Z}_n}(\mathbf{z}) \, \mathrm{d}\mathbf{z}.
\end{align*}
We prove that the first integral on the \gls{rhs} converges to 0 in $\mathbb{P}^\mathbf{Y}$-probability. The other integral is seen to converge using similar arguments.

We have that
\begin{align}
 &\int \left|\int h(\bz') \, \alpha_n(\bz, \bz') \, \varphi(\d\bz'; \bz, \lambda^2 \mathbf{M}_n \mathbf{M}_n^T) - \int h(\bz') \, \alpha(\bz, \bz') \, \varphi(\d\bz'; \bz, \lambda^2 \mathbf{M} \mathbf{M}^T)\right| \, \pi_{\mathbf{Z}_n}(\mathbf{z}) \, \mathrm{d}\mathbf{z} \nonumber \\
 &\quad \leq K \iint \left|\alpha_n(\bz, \bz') \, \varphi(\d\bz'; \bz, \lambda^2 \mathbf{M}_n \mathbf{M}_n^T) - \alpha(\bz, \bz') \, \varphi(\d\bz'; \bz, \lambda^2 \mathbf{M} \mathbf{M}^T)\right| \, \pi_{\mathbf{Z}_n}(\mathbf{z}) \, \mathrm{d}\mathbf{z} \label{eq_ref2} \\
 &\quad \leq K \iint \left|\alpha_n(\bz, \bz') \, \varphi(\d\bz'; \bz, \lambda^2 \mathbf{M}_n \mathbf{M}_n^T) - \alpha_n(\bz, \bz') \, \varphi(\d\bz'; \bz, \lambda^2 \mathbf{M} \mathbf{M}^T)\right| \, \pi_{\mathbf{Z}_n}(\mathbf{z}) \, \mathrm{d}\mathbf{z} \nonumber \\
 &\qquad + K \iint \left|\alpha_n(\bz, \bz') \, \varphi(\d\bz'; \bz, \lambda^2 \mathbf{M} \mathbf{M}^T) - \alpha(\bz, \bz') \, \varphi(\d\bz'; \bz, \lambda^2 \mathbf{M} \mathbf{M}^T)\right| \, \pi_{\mathbf{Z}_n}(\mathbf{z}) \, \mathrm{d}\mathbf{z}, \nonumber
\end{align}
using Jensen's inequality, that there exists a positive constant $K$ such that $|h| \leq K$, and the triangle inequality. We now prove that each of the last two integrals converges to 0. We begin with the first one:
\begin{align*}
 &\iint \left|\alpha_n(\bz, \bz') \, \varphi(\d\bz'; \bz, \lambda^2 \mathbf{M}_n \mathbf{M}_n^T) - \alpha_n(\bz, \bz') \, \varphi(\d\bz'; \bz, \lambda^2 \mathbf{M} \mathbf{M}^T)\right| \, \pi_{\mathbf{Z}_n}(\mathbf{z}) \, \mathrm{d}\mathbf{z} \cr
 &\quad \leq \iint \left|\varphi(\d\bz'; \bz, \lambda^2 \mathbf{M}_n \mathbf{M}_n^T) - \varphi(\d\bz'; \bz, \lambda^2 \mathbf{M} \mathbf{M}^T)\right| \, \pi_{\mathbf{Z}_n}(\mathbf{z}) \, \mathrm{d}\mathbf{z} \cr
 &\quad \leq \left[\text{tr}((\mathbf{M} \mathbf{M}^T)^{-1}\mathbf{M}_n \mathbf{M}_n^T - \mathbf{1}) - \log \det(\mathbf{M}_n \mathbf{M}_n^T (\mathbf{M} \mathbf{M}^T)^{-1}) \right]^{1/2} \rightarrow 0,
\end{align*}
in $\mathbb{P}^\mathbf{Y}$-probability by \autoref{assu:proposal}, using that $0 \leq \alpha_n \leq 1$ and \citet[Proposition 2.1]{devroye2018total}, where $\text{tr}(\, \cdot \,)$ and $\det(\, \cdot \,)$ are the trace and determinant operators, respectively. Note that by \autoref{assu:proposal} we have that $\mathbf{M}_n \mathbf{M}_n^T \rightarrow \mathbf{M} \mathbf{M}^T$ in probability, meaning that all components converge, which implies that the trace and the log of the determinant both vanish.

Next,
\begin{align*}
 &\iint \left|\alpha_n(\bz, \bz') \, \varphi(\d\bz'; \bz, \lambda^2 \mathbf{M} \mathbf{M}^T) - \alpha(\bz, \bz') \, \varphi(\d\bz'; \bz, \lambda^2 \mathbf{M} \mathbf{M}^T)\right| \, \pi_{\mathbf{Z}_n}(\mathbf{z}) \, \mathrm{d}\mathbf{z} \cr
 &\quad \leq \iint \left|\alpha_n(\bz, \bz') \, \pi_{\mathbf{Z}_n}(\mathbf{z})  - \alpha(\bz, \bz') \, \varphi(\bz; \mathbf{0}, \iF{\boldsymbol\theta_0})\right| \, \varphi(\d\bz'; \bz, \lambda^2 \mathbf{M} \mathbf{M}^T)  \, \mathrm{d}\mathbf{z} \cr
 &\qquad + \iint \left|\alpha(\bz, \bz') \, \varphi(\bz; \mathbf{0}, \iF{\boldsymbol\theta_0})  - \alpha(\bz, \bz') \, \pi_{\mathbf{Z}_n}(\mathbf{z})\right| \, \varphi(\d\bz'; \bz, \lambda^2 \mathbf{M} \mathbf{M}^T)  \, \mathrm{d}\mathbf{z},
\end{align*}
using the triangle inequality. The second integral is seen to converge to 0 because
\begin{align}\label{eq_ref1}
 \hspace{-5mm}&\iint \left|\alpha(\bz, \bz') \, \varphi(\bz; \mathbf{0}, \iF{\boldsymbol\theta_0})  - \alpha(\bz, \bz') \, \pi_{\mathbf{Z}_n}(\mathbf{z})\right| \, \varphi(\d\bz'; \bz, \lambda^2 \mathbf{M} \mathbf{M}^T)  \, \mathrm{d}\mathbf{z} \\
 &\leq \int \left|\varphi(\bz; \mathbf{0}, \iF{\boldsymbol\theta_0})  -  \pi_{\mathbf{Z}_n}(\mathbf{z})\right| \, \mathrm{d}\mathbf{z} \nonumber \\
 &= \int\left|\varphi(\boldsymbol\theta; \hat{\boldsymbol\theta}_{n}, \iF{\boldsymbol\theta_0}/n) - {\pi}_{n}(\boldsymbol\theta)\right| \mathrm{d}\boldsymbol\theta \rightarrow 0 \nonumber
\end{align}
in $\mathbb{P}^\mathbf{Y}$-probability by \autoref{assu:posterior}, using that $0 \leq \alpha \leq 1$ and a change of variable $\boldsymbol\theta = \mathbf{z} / n^{1/2} +  \hat{\boldsymbol\theta}_n$. For the first integral, we write
\begin{align*}
 &\iint \left|\alpha_n(\bz, \bz') \, \pi_{\mathbf{Z}_n}(\mathbf{z})  - \alpha(\bz, \bz') \, \varphi(\bz; \mathbf{0}, \iF{\boldsymbol\theta_0})\right| \, \varphi(\d\bz'; \bz, \lambda^2 \mathbf{M} \mathbf{M}^T)  \, \mathrm{d}\mathbf{z} \cr
 &\quad = \iint \left|\min\{\pi_{\mathbf{Z}_n}(\mathbf{z}), \pi_{\mathbf{Z}_n}(\mathbf{z}')\}   - \min\{\varphi(\bz; \mathbf{0},\iF{\boldsymbol\theta_0}), \varphi(\bz'; \mathbf{0}, \iF{\boldsymbol\theta_0})\}\right| \, \varphi(\d\bz'; \bz, \lambda^2 \mathbf{M} \mathbf{M}^T)  \, \mathrm{d}\mathbf{z} \cr
 &\quad \leq \iint \left|\pi_{\mathbf{Z}_n}(\mathbf{z})   - \varphi(\bz; \mathbf{0},\iF{\boldsymbol\theta_0})\right| \, \varphi(\d\bz'; \bz, \lambda^2 \mathbf{M} \mathbf{M}^T)  \, \mathrm{d}\mathbf{z} \cr
 &\qquad + \iint \left|\pi_{\mathbf{Z}_n}(\mathbf{z}') - \varphi(\bz'; \mathbf{0}, \iF{\boldsymbol\theta_0})\right| \, \varphi(\d\bz'; \bz, \lambda^2 \mathbf{M} \mathbf{M}^T)  \, \mathrm{d}\mathbf{z},
\end{align*}
using that $|\min\{a, b\} - \min\{c, d\}| \leq |a - c| + |b - d|$ for any real numbers $a, b, c$ and $d$. It is seen that both integrals on the \gls{rhs} vanish as above (recall \eqref{eq_ref1}) after noticing that $\varphi(\d\bz'; \bz, \lambda^2 \mathbf{M} \mathbf{M}^T) \, \mathrm{d}\mathbf{z} = \varphi(\d\bz; \bz', \lambda^2 \mathbf{M} \mathbf{M}^T) \, \mathrm{d}\mathbf{z}'$, which is used in the second integral.

There remains to verify Condition 3: the continuity of $Ph$. Without loss of generality, consider a non-random sequence of vectors $(\mathbf{e}_n)_{n \geq 1}$ with monotonically shrinking components (in absolute value) such that $\sup_n \mathbf{e}_n^T (\mathbf{M} \mathbf{M}^T)^{-1} \mathbf{e}_n < \infty$. We now prove that $Ph(\bz + \mathbf{e}_n) \rightarrow Ph(\bz)$ as $n \rightarrow \infty$.

We have that
\[
 Ph(\mathbf{z} + \mathbf{e}_n) = \int h(\bz') \, \alpha(\bz + \mathbf{e}_n, \bz') \, \varphi(\d\bz'; \bz + \mathbf{e}_n, \lambda^2 \mathbf{M} \mathbf{M}^T) + h(\bz + \mathbf{e}_n) \, \rho(\bz + \mathbf{e}_n).
\]
We prove that the first term on the \gls{rhs} converges to
\[
 \int h(\bz') \, \alpha(\bz, \bz') \, \varphi(\d\bz'; \bz, \lambda^2 \mathbf{M} \mathbf{M}^T);
\]
the convergence of the second term follows using similar arguments.

We write
\begin{align*}
 &\int h(\bz') \, \alpha(\bz + \mathbf{e}_n, \bz') \, \varphi(\d\bz'; \bz + \mathbf{e}_n, \lambda^2 \mathbf{M} \mathbf{M}^T) \cr
  &\hspace{-7mm} = \exp\left\{-\frac{\mathbf{e}_n^T (\lambda^2 \mathbf{M} \mathbf{M}^T)^{-1} \mathbf{e}_n}{2}\right\} \int h(\bz') \, \alpha(\bz + \mathbf{e}_n, \bz') \, \exp\left\{-\mathbf{e}_n^T (\lambda^2 \mathbf{M} \mathbf{M}^T)^{-1} (\bz' - \bz)\right\} \, \varphi(\d\bz'; \bz, \lambda^2 \mathbf{M} \mathbf{M}^T) \cr
  &\hspace{-7mm}=\exp\left\{-\frac{\mathbf{e}_n^T (\lambda^2 \mathbf{M} \mathbf{M}^T)^{-1} \mathbf{e}_n}{2}\right\} \E_{\bz}\left[h(\mathbf{Z}') \, \alpha(\bz + \mathbf{e}_n, \mathbf{Z}') \, \exp\left\{-\mathbf{e}_n^T (\lambda^2 \mathbf{M} \mathbf{M}^T)^{-1} (\mathbf{Z}' - \bz)\right\}\right],
\end{align*}
where the expectation is with respect to $\varphi(\, \cdot \,; \bz, \lambda^2 \mathbf{M} \mathbf{M}^T)$; we highlight a dependence on $\bz$ using the notation $\E_{\bz}$.

We have that
\[
 \exp\left\{-\frac{\mathbf{e}_n^T (\lambda^2 \mathbf{M} \mathbf{M}^T)^{-1} \mathbf{e}_n}{2}\right\} \rightarrow 1,
\]
and
\[
 h(\mathbf{Z}') \, \alpha(\bz + \mathbf{e}_n, \mathbf{Z}') \, \exp\left\{-\mathbf{e}_n^T (\lambda^2 \mathbf{M} \mathbf{M}^T)^{-1} (\mathbf{Z}' - \bz)\right\} \rightarrow h(\mathbf{Z}') \, \alpha(\bz, \mathbf{Z}'),
\]
almost surely, given the continuity of $\alpha$ and the exponential function.

To prove that the expectation converges to
\[
 \E_{\bz}\left[h(\mathbf{Z}') \, \alpha(\bz, \mathbf{Z}')\right] = \int h(\bz') \, \alpha(\bz, \bz') \, \varphi(\d\bz'; \bz, \lambda^2 \mathbf{M} \mathbf{M}^T),
 \]
we thus only need to prove that
\[
h(\mathbf{Z}') \, \alpha(\bz + \mathbf{e}_n, \mathbf{Z}') \, \exp\left\{-\mathbf{e}_n^T (\lambda^2 \mathbf{M} \mathbf{M}^T)^{-1} (\mathbf{Z}' - \bz)\right\}
\]
is uniformly integrable. To prove this, we show that
\[
 \sup_n \E\left[\left(h(\mathbf{Z}') \, \alpha(\bz + \mathbf{e}_n, \mathbf{Z}') \, \exp\left\{-\mathbf{e}_n^T (\lambda^2 \mathbf{M} \mathbf{M}^T)^{-1} (\mathbf{Z}' - \bz)\right\}\right)^2\right] < \infty.
\]

We have that
\begin{align*}
& \E\left[\left(h(\mathbf{Z}') \, \alpha(\bz + \mathbf{e}_n, \mathbf{Z}') \, \exp\left\{-\mathbf{e}_n^T (\lambda^2 \mathbf{M} \mathbf{M}^T)^{-1} (\mathbf{Z}' - \bz)\right\}\right)^2\right] \cr
&\qquad \leq K^2 \E\left[\exp\left\{-2\,\mathbf{e}_n^T (\lambda^2 \mathbf{M} \mathbf{M}^T)^{-1} (\mathbf{Z}' - \bz)\right\}\right] = K^2 \exp\left\{2 \, \lambda^{-2} \, \mathbf{e}_n^T (\mathbf{M} \mathbf{M}^T)^{-1} \mathbf{e}_n\right\}.
\end{align*}
This concludes the proof of Result (i).

\emph{Result (ii).}  We want to prove that
\begin{align*}
 \left|\iint \alpha_n(\bz, \bz') \, \varphi(\d\bz'; \bz, \lambda^2 \mathbf{M}_n \mathbf{M}_n^T) \, \pi_{\mathbf{Z}_n}(\mathbf{z}) \, \mathrm{d}\mathbf{z} - \iint \alpha(\bz, \bz') \, \varphi(\d\bz'; \bz, \lambda^2 \mathbf{M} \mathbf{M}^T) \, \varphi(\d\bz; \mathbf{0}, \iF{\boldsymbol\theta_0})\right| \rightarrow 0,
\end{align*}
in $\mathbb{P}^\mathbf{Y}$-probability as $n \rightarrow \infty$. Using the triangle and Jensen's inequality and that $0 \leq \alpha \leq 1$,
\begin{align*}
 &\left|\iint \alpha_n(\bz, \bz') \, \varphi(\d\bz'; \bz, \lambda^2 \mathbf{M}_n \mathbf{M}_n^T) \, \pi_{\mathbf{Z}_n}(\mathbf{z}) \, \mathrm{d}\mathbf{z} - \iint \alpha(\bz, \bz') \, \varphi(\d\bz'; \bz, \lambda^2 \mathbf{M} \mathbf{M}^T) \, \varphi(\d\bz; \mathbf{0}, \iF{\boldsymbol\theta_0})\right| \cr
 &\quad \leq \left|\iint \alpha_n(\bz, \bz') \, \varphi(\d\bz'; \bz, \lambda^2 \mathbf{M}_n \mathbf{M}_n^T) \, \pi_{\mathbf{Z}_n}(\mathbf{z}) \, \mathrm{d}\mathbf{z} - \iint \alpha(\bz, \bz') \, \varphi(\d\bz'; \bz, \lambda^2 \mathbf{M} \mathbf{M}^T) \,  \pi_{\mathbf{Z}_n}(\mathbf{z}) \, \mathrm{d}\mathbf{z}\right| \cr
 &\qquad + \left|\iint \alpha(\bz, \bz') \, \varphi(\d\bz'; \bz, \lambda^2 \mathbf{M} \mathbf{M}^T) \,  \pi_{\mathbf{Z}_n}(\mathbf{z}) \, \mathrm{d}\mathbf{z} - \iint \alpha(\bz, \bz') \, \varphi(\d\bz'; \bz, \lambda^2 \mathbf{M} \mathbf{M}^T) \,  \varphi(\d\bz; \mathbf{0}, \iF{\boldsymbol\theta_0})\right| \cr
 &\quad \leq \iint \left|\alpha_n(\bz, \bz') \, \varphi(\d\bz'; \bz, \lambda^2 \mathbf{M}_n \mathbf{M}_n^T) - \alpha(\bz, \bz') \, \varphi(\d\bz'; \bz, \lambda^2 \mathbf{M} \mathbf{M}^T) \right| \,  \pi_{\mathbf{Z}_n}(\mathbf{z}) \, \mathrm{d}\mathbf{z} \cr
 &\qquad + \int \left| \pi_{\mathbf{Z}_n}(\mathbf{z}) - \varphi(\bz; \mathbf{0}, \iF{\boldsymbol\theta_0})\right| \, \mathrm{d}\mathbf{z}.
\end{align*}
We shown in the proof of Result (i) that both integrals converge to 0 (recall \eqref{eq_ref2} and \eqref{eq_ref1}), which concludes the proof of Result (ii).

\emph{Result (iii).} To prove this result, we show that
$$\E\left[\|\lambda \mathbf{M}_n \boldsymbol\epsilon\|_{\F{\hat{\boldsymbol\theta}_n}}^2 \alpha_n(\mathbf{Z}_n, \mathbf{Z}_n + \lambda \mathbf{M}_n \boldsymbol\epsilon)\right] -  \E\left[\|\lambda \mathbf{M} \boldsymbol\epsilon\|_{\F{\boldsymbol\theta_0}}^2 \alpha(\mathbf{Z}, \mathbf{Z} + \lambda \mathbf{M} \boldsymbol\epsilon)\right]  \rightarrow 0,$$
in $\mathbb{P}^\mathbf{Y}$-probability, where $\mathbf{Z}_n \sim \pi_{\mathbf{Z}_n}$ and $\mathbf{Z} \sim \varphi(\, \cdot \,; \mathbf{0}, \iF{\boldsymbol\theta_0})$, under the assumption that
\[
 \left|\E\left[\|\lambda \mathbf{M}_n \boldsymbol\epsilon\|_{\F{\hat{\boldsymbol\theta}_n}}^2 \alpha_n(\mathbf{Z}_n, \mathbf{Z}_n + \lambda \mathbf{M}_n \boldsymbol\epsilon)\right] -  \E\left[\|\lambda \mathbf{M}\boldsymbol\epsilon\|_{\F{\boldsymbol\theta_0}}^2 \alpha_n(\mathbf{Z}_n, \mathbf{Z}_n + \lambda \mathbf{M}\boldsymbol\epsilon)\right]\right| \rightarrow 0,
\]
which will be seen to imply Result (iii). Indeed, this assumption is more general than $\mathbf{M}_n \mathbf{M}_n^T = \iF{\hat{\boldsymbol\theta}_n} \rightarrow \mathbf{M}\mathbf{M}^T = \iF{\boldsymbol\theta_0}$; we will show below that it is verified when $\mathbf{M}_n \mathbf{M}_n^T = \iF{\hat{\boldsymbol\theta}_n} \rightarrow \mathbf{M}\mathbf{M}^T = \iF{\boldsymbol\theta_0}$.

Using the triangle inequality,
\begin{align*}
 &\left|\E\left[\|\lambda \mathbf{M}_n \boldsymbol\epsilon\|_{\F{\hat{\boldsymbol\theta}_n}}^2 \alpha_n(\mathbf{Z}_n, \mathbf{Z}_n + \lambda \mathbf{M}_n \boldsymbol\epsilon)\right] -  \E\left[\|\lambda \mathbf{M} \boldsymbol\epsilon\|_{\F{\boldsymbol\theta_0}}^2 \alpha(\mathbf{Z}, \mathbf{Z} + \lambda \mathbf{M} \boldsymbol\epsilon)\right]\right| \cr
 &\quad \leq  \left|\E\left[\|\lambda \mathbf{M}_n \boldsymbol\epsilon\|_{\F{\hat{\boldsymbol\theta}_n}}^2 \alpha_n(\mathbf{Z}_n, \mathbf{Z}_n + \lambda \mathbf{M}_n \boldsymbol\epsilon)\right] -  \E\left[\|\lambda \mathbf{M}\boldsymbol\epsilon\|_{\F{\boldsymbol\theta_0}}^2 \alpha_n(\mathbf{Z}_n, \mathbf{Z}_n + \lambda \mathbf{M}\boldsymbol\epsilon)\right]\right| \cr
 &\qquad +  \left|\E\left[\|\lambda \mathbf{M}\boldsymbol\epsilon\|_{\F{\boldsymbol\theta_0}}^2 \alpha_n(\mathbf{Z}_n, \mathbf{Z}_n + \lambda \mathbf{M}\boldsymbol\epsilon)\right] - \E\left[\|\lambda \mathbf{M} \boldsymbol\epsilon\|_{\F{\boldsymbol\theta_0}}^2 \alpha(\mathbf{Z}, \mathbf{Z} + \lambda \mathbf{M} \boldsymbol\epsilon)\right]\right|.
\end{align*}

The first absolute value on the \gls{rhs} vanishes by assumption. We now prove that the second absolute value on the \gls{rhs} vanishes. We have,
 \begin{align*}
  &\left|\E\left[\|\lambda \mathbf{M} \boldsymbol\epsilon\|_{\F{\boldsymbol\theta_0}}^2 \alpha_n(\mathbf{Z}_n, \mathbf{Z}_n + \lambda \mathbf{M} \boldsymbol\epsilon)\right] - \E\left[\|\lambda \mathbf{M} \boldsymbol\epsilon\|_{\F{\boldsymbol\theta_0}}^2 \alpha(\mathbf{Z}, \mathbf{Z} + \lambda \mathbf{M} \boldsymbol\epsilon)\right]\right| \cr
  &\quad = \left|\iint \|\lambda \mathbf{M} \boldsymbol\epsilon\|_{\F{\boldsymbol\theta_0}}^2 \min\{\pi_{\mathbf{Z}_n}(\mathbf{z}), \pi_{\mathbf{Z}_n}(\mathbf{z} + \lambda \mathbf{M} \boldsymbol\epsilon)\}  \, \varphi(\d\boldsymbol\epsilon; \mathbf{0}, \mathbf{1})  \, \mathrm{d}\mathbf{z} \right. \cr
   &\qquad \left.- \iint \|\lambda \mathbf{M} \boldsymbol\epsilon\|_{\F{\boldsymbol\theta_0}}^2 \min\{\varphi(\bz; \mathbf{0}, \iF{\boldsymbol\theta_0}), \varphi(\bz + \lambda \mathbf{M} \boldsymbol\epsilon; \mathbf{0}, \iF{\boldsymbol\theta_0})\}  \, \varphi(\d\boldsymbol\epsilon; \mathbf{0}, \mathbf{1})  \, \mathrm{d}\mathbf{z} \right| \cr
   &\quad \leq \iint \|\lambda \mathbf{M} \boldsymbol\epsilon\|_{\F{\boldsymbol\theta_0}}^2 \left|\min\{\pi_{\mathbf{Z}_n}(\mathbf{z}), \pi_{\mathbf{Z}_n}(\mathbf{z} + \lambda \mathbf{M} \boldsymbol\epsilon)\} \right. \cr
    &\qquad \left.- \min\{\varphi(\bz; \mathbf{0}, \iF{\boldsymbol\theta_0}), \varphi(\bz + \lambda \mathbf{M} \boldsymbol\epsilon; \mathbf{0}, \iF{\boldsymbol\theta_0})\} \right| \, \varphi(\d\boldsymbol\epsilon; \mathbf{0}, \mathbf{1})  \, \mathrm{d}\mathbf{z} \cr
    &\quad \leq \iint \|\lambda \mathbf{M} \boldsymbol\epsilon\|_{\F{\boldsymbol\theta_0}}^2 \left|\pi_{\mathbf{Z}_n}(\mathbf{z}) - \varphi(\bz; \mathbf{0}, \iF{\boldsymbol\theta_0})\right| \, \varphi(\d\boldsymbol\epsilon; \mathbf{0}, \mathbf{1})  \, \mathrm{d}\mathbf{z} \cr
    &\qquad + \iint \|\lambda \mathbf{M} \boldsymbol\epsilon\|_{\F{\boldsymbol\theta_0}}^2 \left|\pi_{\mathbf{Z}_n}(\mathbf{z} + \lambda \mathbf{M} \boldsymbol\epsilon) - \varphi(\bz + \lambda \mathbf{M} \boldsymbol\epsilon; \mathbf{0}, \iF{\boldsymbol\theta_0})\right| \, \varphi(\d\boldsymbol\epsilon; \mathbf{0}, \mathbf{1})  \, \mathrm{d}\mathbf{z},
 \end{align*}
 using Jensen's inequality and $|\min\{a, b\} - \min\{c, d\}| \leq |a - c| + |b - d|$ for any real numbers $a, b, c$ and $d$.

The first integral on the \gls{rhs} vanishes for the same reasons we have seen before (recall \eqref{eq_ref1}). We rewrite the second one as:
\begin{align}\label{eq_ref3}
 &\iint \|\bz' - \bz\|_{\F{\boldsymbol\theta_0}}^2 \left|\pi_{\mathbf{Z}_n}(\bz') - \varphi(\bz'; \mathbf{0}, \iF{\boldsymbol\theta_0})\right| \, \varphi(\d\bz'; \bz, \lambda^2 \mathbf{M} \mathbf{M}^T )  \, \mathrm{d}\mathbf{z} \nonumber \\
 &\quad = \iint \|\lambda\boldsymbol\epsilon\|^2 \left|\pi_{\mathbf{Z}_n}(\bz') - \varphi(\bz'; \mathbf{0}, \iF{\boldsymbol\theta_0})\right| \, \varphi(\d\boldsymbol\epsilon; \mathbf{0}, \mathbf{1})  \, \mathrm{d}\mathbf{z}',
\end{align}
using that $\varphi(\d\bz'; \bz, \lambda^2 \mathbf{M} \mathbf{M}^T ) \, \mathrm{d}\mathbf{z} = \varphi(\d\bz; \bz', \lambda^2 \mathbf{M} \mathbf{M}^T ) \, \mathrm{d}\mathbf{z}'$ and a change of variables $\boldsymbol\epsilon = (\lambda \mathbf{M})^{-1}(\bz - \bz')$. The last integral vanishes as seen before (recall \eqref{eq_ref1}).

We finish the proof by showing that the assumption
\[
  \left|\E\left[\|\lambda\mathbf{M}_n \boldsymbol\epsilon\|_{\F{\hat{\boldsymbol\theta}_n}}^2 \alpha_n(\mathbf{Z}_n, \mathbf{Z}_n + \lambda \mathbf{M}_n \boldsymbol\epsilon)\right] -  \E\left[\|\lambda \mathbf{M}\boldsymbol\epsilon\|_{\F{\boldsymbol\theta_0}}^2 \alpha_n(\mathbf{Z}_n, \mathbf{Z}_n + \lambda \mathbf{M}\boldsymbol\epsilon)\right]\right| \rightarrow 0,
\]
is verified when $\mathbf{M}_n \mathbf{M}_n^T = \iF{\hat{\boldsymbol\theta}_n} \rightarrow \mathbf{M}\mathbf{M}^T = \iF{\boldsymbol\theta_0}$. In this case,
\begin{align*}
 &\left|\E\left[\|\lambda\mathbf{M}_n \boldsymbol\epsilon\|_{\F{\hat{\boldsymbol\theta}_n}}^2 \alpha_n(\mathbf{Z}_n, \mathbf{Z}_n + \lambda \mathbf{M}_n \boldsymbol\epsilon)\right] -  \E\left[\|\lambda \mathbf{M}\boldsymbol\epsilon\|_{\F{\boldsymbol\theta_0}}^2 \alpha_n(\mathbf{Z}_n, \mathbf{Z}_n + \lambda \mathbf{M}\boldsymbol\epsilon)\right]\right| \cr
 &\quad = \left|\E\left[\|\lambda\boldsymbol\epsilon\|^2 \alpha_n(\mathbf{Z}_n, \mathbf{Z}_n + \lambda \mathbf{M}_n \boldsymbol\epsilon)\right] -  \E\left[\|\lambda \boldsymbol\epsilon\|^2 \alpha_n(\mathbf{Z}_n, \mathbf{Z}_n + \lambda \mathbf{M}\boldsymbol\epsilon)\right]\right| \cr
 &\quad = \left| \iint\|\lambda\boldsymbol\epsilon\|^2  \min\{\pi_{\mathbf{Z}_n}(\mathbf{z}), \pi_{\mathbf{Z}_n}(\mathbf{z} + \lambda \mathbf{M}_n \boldsymbol\epsilon)\} \, \varphi(\d\boldsymbol\epsilon; \mathbf{0}, \mathbf{1})  \, \mathrm{d}\mathbf{z} \right. \cr
 &\qquad - \left. \iint\|\lambda\boldsymbol\epsilon\|^2  \min\{\pi_{\mathbf{Z}_n}(\mathbf{z}), \pi_{\mathbf{Z}_n}(\mathbf{z} + \lambda \mathbf{M} \boldsymbol\epsilon)\} \, \varphi(\d\boldsymbol\epsilon; \mathbf{0}, \mathbf{1})  \, \mathrm{d}\mathbf{z} \right| \cr
 &\quad \leq  \iint\|\lambda\boldsymbol\epsilon\|^2 \left| \min\{\pi_{\mathbf{Z}_n}(\mathbf{z}), \pi_{\mathbf{Z}_n}(\mathbf{z} + \lambda \mathbf{M}_n \boldsymbol\epsilon)\} - \min\{\pi_{\mathbf{Z}_n}(\mathbf{z}), \pi_{\mathbf{Z}_n}(\mathbf{z} + \lambda \mathbf{M} \boldsymbol\epsilon)\} \right| \, \varphi(\d\boldsymbol\epsilon; \mathbf{0}, \mathbf{1})  \, \mathrm{d}\mathbf{z} \cr
 &\quad \leq \iint\|\lambda\boldsymbol\epsilon\|^2 \left| \pi_{\mathbf{Z}_n}(\mathbf{z} + \lambda \mathbf{M}_n \boldsymbol\epsilon) - \pi_{\mathbf{Z}_n}(\mathbf{z} + \lambda \mathbf{M} \boldsymbol\epsilon) \right| \, \varphi(\d\boldsymbol\epsilon; \mathbf{0}, \mathbf{1})  \, \mathrm{d}\mathbf{z},
\end{align*}
using Jensen's inequality and $|\min\{a, b\} - \min\{c, d\}| \leq |a - c| + |b - d|$ for any real numbers $a, b, c$ and $d$.

Now, using the triangle inequality,
\begin{align}\label{eqn_ine_result_iii}
& \iint\|\lambda\boldsymbol\epsilon\|^2 \left| \pi_{\mathbf{Z}_n}(\mathbf{z} + \lambda \mathbf{M}_n \boldsymbol\epsilon) - \pi_{\mathbf{Z}_n}(\mathbf{z} + \lambda \mathbf{M} \boldsymbol\epsilon) \right| \, \varphi(\d\boldsymbol\epsilon; \mathbf{0}, \mathbf{1})  \, \mathrm{d}\mathbf{z} \cr
&\quad \leq \iint\|\lambda\boldsymbol\epsilon\|^2 \left| \pi_{\mathbf{Z}_n}(\mathbf{z} + \lambda \mathbf{M}_n \boldsymbol\epsilon) -  \varphi(\bz + \lambda \mathbf{M}_n \boldsymbol\epsilon; \mathbf{0}, \iF{\boldsymbol\theta_0}) \right| \, \varphi(\d\boldsymbol\epsilon; \mathbf{0}, \mathbf{1})  \, \mathrm{d}\mathbf{z} \cr
&\qquad + \iint\|\lambda\boldsymbol\epsilon\|^2 \left| \varphi(\bz + \lambda \mathbf{M}_n \boldsymbol\epsilon; \mathbf{0}, \iF{\boldsymbol\theta_0}) -  \varphi(\bz + \lambda \mathbf{M} \boldsymbol\epsilon; \mathbf{0}, \iF{\boldsymbol\theta_0})\right| \, \varphi(\d\boldsymbol\epsilon; \mathbf{0}, \mathbf{1})  \, \mathrm{d}\mathbf{z} \cr
&\qquad + \iint\|\lambda\boldsymbol\epsilon\|^2 \left| \varphi(\bz + \lambda \mathbf{M} \boldsymbol\epsilon; \mathbf{0}, \iF{\boldsymbol\theta_0}) - \pi_{\mathbf{Z}_n}(\mathbf{z} + \lambda \mathbf{M} \boldsymbol\epsilon)\right| \, \varphi(\d\boldsymbol\epsilon; \mathbf{0}, \mathbf{1})  \, \mathrm{d}\mathbf{z}.
\end{align}

We now prove that each of the integrals on the \gls{rhs} vanishes. We start with the first one,
\begin{align*}
 &\iint\|\lambda\boldsymbol\epsilon\|^2 \left| \pi_{\mathbf{Z}_n}(\mathbf{z} + \lambda \mathbf{M}_n \boldsymbol\epsilon) -  \varphi(\bz + \lambda \mathbf{M}_n \boldsymbol\epsilon; \mathbf{0}, \iF{\boldsymbol\theta_0}) \right| \, \varphi(\d\boldsymbol\epsilon; \mathbf{0}, \mathbf{1})  \, \mathrm{d}\mathbf{z} \cr
 &\quad = \iint\|\bz' - \bz\|_{\F{\hat{\boldsymbol\theta}_n}}^2 \left| \pi_{\mathbf{Z}_n}(\bz') -  \varphi(\bz'; \mathbf{0}, \iF{\boldsymbol\theta_0}) \right| \, \varphi(\d\bz'; \bz, \lambda^2 \mathbf{M}_n \mathbf{M}_n^T)  \, \mathrm{d}\mathbf{z},
\end{align*}
using the change of variable $\bz' = \mathbf{z} + \lambda \mathbf{M}_n \boldsymbol\epsilon$. As we have seen before, the last integral vanishes (recall \eqref{eq_ref3}). The third integral on the \gls{rhs} in \eqref{eqn_ine_result_iii} vanishes for similar reasons.

For the second one, we use that $\mathbf{M}_n \rightarrow \mathbf{M}$ in $\mathbb{P}^\mathbf{Y}$-probability. This is true because $\mathbf{M}_n \mathbf{M}_n^T \rightarrow \mathbf{M} \mathbf{M}^T$ in $\mathbb{P}^\mathbf{Y}$-probability and the Cholesky decomposition yields a continuous map. Now, using \citet[Proposition 2.1]{devroye2018total} and Cauchy--Schwarz inequality,
\begin{align*}
 &\iint\|\lambda\boldsymbol\epsilon\|^2 \left| \varphi(\bz; -\lambda \mathbf{M}_n \boldsymbol\epsilon, \iF{\boldsymbol\theta_0}) -  \varphi(\bz; -\lambda \mathbf{M} \boldsymbol\epsilon, \iF{\boldsymbol\theta_0})\right| \, \varphi(\d\boldsymbol\epsilon; \mathbf{0}, \mathbf{1})  \, \mathrm{d}\mathbf{z} \cr
 &\quad \leq \int\|\lambda\boldsymbol\epsilon\|^2 \lambda \left[\boldsymbol\epsilon^T (\mathbf{M}^{-1} \mathbf{M}_n - \mathbf{1})^T  (\mathbf{M}^{-1} \mathbf{M}_n - \mathbf{1})\boldsymbol\epsilon\right]^{1/2}   \, \varphi(\d\boldsymbol\epsilon; \mathbf{0}, \mathbf{1})  \cr
 &\quad \leq \lambda^3 \left[\int \|\epsilon\|^4 \, \varphi(\d\boldsymbol\epsilon; \mathbf{0}, \mathbf{1})\right]^{1/2} \left[\int \boldsymbol\epsilon^T (\mathbf{M}^{-1} \mathbf{M}_n - \mathbf{1})^T  (\mathbf{M}^{-1} \mathbf{M}_n - \mathbf{1})\boldsymbol\epsilon \, \varphi(\d\boldsymbol\epsilon; \mathbf{0}, \mathbf{1})\right]^{1/2}.
\end{align*}
The first integral on the \gls{rhs} is bounded. We write the second one as an expectation:
\begin{align*}
 \E[\boldsymbol\epsilon^T \mathbf{A}_n \boldsymbol\epsilon] = \E[\boldsymbol\epsilon^T \mathbf{Q}_n \boldsymbol\Lambda_n \mathbf{Q}_n^T \boldsymbol\epsilon] = \E\left[\sum_{j = 1}^d \lambda_{j, n} \xi_{j, n}^2\right] = \sum_{j = 1}^d \lambda_{j, n} = \text{tr}(\mathbf{A}_n) \rightarrow 0,
\end{align*}
using an eigendecomposition of $\mathbf{A}_n$ and that $\boldsymbol\xi_n := (\xi_{1, n}, \ldots, \xi_{d, n})^T := \mathbf{Q}_n^T \boldsymbol\epsilon$ is a random vector with independent standard normal components, where $\mathbf{A}_n := (\mathbf{M}^{-1} \mathbf{M}_n - \mathbf{1})^T  (\mathbf{M}^{-1} \mathbf{M}_n - \mathbf{1})$, $\mathbf{Q}_n$ is an orthogonal matrix whose columns are the eigenvectors of $\mathbf{A}_n$, and $\boldsymbol\Lambda_n$ is a diagonal matrix whose entries $\lambda_{1, n}, \ldots, \lambda_{d, n}$ are the eigenvalues of $\mathbf{A}_n$. This concludes the proof.
\end{proof}

\begin{proof}[Proof of \autoref{cor:esjd}]
 We first denote $\mathbf{S} := \lambda \mathbf{M}$ and thus note that $\mathbf{Z}' = \mathbf{Z} + \mathbf{S}\boldsymbol\epsilon$, where $\mathbf{Z} \sim \varphi(\, \cdot \,; \mathbf{0}, \iF{\boldsymbol\theta_0})$ and $\boldsymbol\epsilon \sim \varphi(\, \cdot \,; \mathbf{0}, \mathbf{1})$. We have
\begin{align*}
	\textsc{esjd}(\lambda, \mathbf{M}) &= \E\left[\|\mathbf{S}\boldsymbol{\epsilon}\|_{\F{\boldsymbol\theta_0}}^2 \min\left\{1, \frac{\varphi(\mathbf{Z} + \mathbf{S}\boldsymbol\epsilon; \mathbf{0}, \iF{\boldsymbol\theta_0})}{\varphi(\mathbf{Z}; \mathbf{0}, \iF{\boldsymbol\theta_0})} \right\} \right] \\
	&= \E\left[\|\mathbf{S}\boldsymbol{\epsilon}\|_{\F{\boldsymbol\theta_0}}^2 \min\left\{1, \exp\left(- \frac{1}{2}\boldsymbol{\epsilon}^T \mathbf{S}^T \F{\boldsymbol\theta_0} \, \mathbf{S} \boldsymbol{\epsilon} +  \mathbf{Z}^T \mathbf{S}^T \F{\boldsymbol\theta_0} \, \boldsymbol{\epsilon}\right)\right\} \right] \\
	&= \E\left[\|\mathbf{S}\boldsymbol{\epsilon}\|_{\F{\boldsymbol\theta_0}}^2  \E\left[\min\left\{1, \exp\left(- \frac{1}{2}\boldsymbol{\epsilon}^T \mathbf{S}^T \F{\boldsymbol\theta_0} \, \mathbf{S} \boldsymbol{\epsilon} +  \mathbf{Z}^T \mathbf{S}^T \F{\boldsymbol\theta_0} \, \boldsymbol{\epsilon}\right)\right\} \mid \boldsymbol{\epsilon}\right] \right].
\end{align*}

In the following, we make use of the fact that for a univariate normal random variable $X$ with $X \sim \varphi(\,\cdot\,; m, s^2)$, we have that
\begin{equation*}
	\E\left[\min\{1, \exp(X)\}\right] = \Phi\left(\frac{m}{s}\right) + \exp\left(m + \frac{s^2}{2}\right) \, \Phi\left(-s - \frac{m}{s}\right).
\end{equation*}
In particular, if $m=-s^2/2$,
\begin{equation}\label{eq:min_one_gaussian}
	\E\left[\min\{1, \exp(X)\}\right] = 2\Phi\left(-\frac{s}{2}\right).
\end{equation}

Consider the case where $\mathbf{M} = \mathbf{1}$. Thus, given $\boldsymbol{\epsilon}$, $- \frac{1}{2}\boldsymbol{\epsilon}^T \mathbf{S}^T \F{\boldsymbol\theta_0} \, \mathbf{S} \boldsymbol{\epsilon} +  \mathbf{Z}^T \mathbf{S}^T \F{\boldsymbol\theta_0} \, \boldsymbol{\epsilon} = - \frac{\lambda^2}{2}\boldsymbol{\epsilon}^T \F{\boldsymbol\theta_0} \, \boldsymbol{\epsilon} +  \lambda \mathbf{Z}^T \F{\boldsymbol\theta_0} \, \boldsymbol{\epsilon}$ is a Gaussian random variable with $m = - \frac{\lambda^2}{2}\boldsymbol{\epsilon}^T \F{\boldsymbol\theta_0} \, \boldsymbol{\epsilon}$ and $s^2 = \lambda^2 \boldsymbol{\epsilon}^T \F{\boldsymbol\theta_0} \, \boldsymbol{\epsilon}$, implying that
\begin{align*}
 &\E\left[\|\mathbf{S}\boldsymbol{\epsilon}\|_{\F{\boldsymbol\theta_0}}^2  \E\left[\min\left\{1, \exp\left(- \frac{1}{2}\boldsymbol{\epsilon}^T \mathbf{S}^T \F{\boldsymbol\theta_0} \, \mathbf{S} \boldsymbol{\epsilon} +  \mathbf{Z}^T \mathbf{S}^T \F{\boldsymbol\theta_0} \, \boldsymbol{\epsilon}\right)\right\} \mid \boldsymbol{\epsilon}\right] \right] \cr
 &\qquad = 2 \lambda^2 \E\left[\|\boldsymbol{\epsilon}\|_{\F{\boldsymbol\theta_0}}^2  \Phi\left(-\lambda \, \frac{\|\boldsymbol{\epsilon}\|_{\F{\boldsymbol\theta_0}}}{2}\right) \right].
\end{align*}

The formulae for \textsc{esjd} with $\mathbf{M}$ such that $\mathbf{M}\mathbf{M}^T = \mathcal{I}_{\boldsymbol\theta_0}^{-1}$ and the expected acceptance probabilities are derived analogously.
\end{proof}

\begin{proof}[Proof of \autoref{prop:decreasing}]
The result follows directly from a result in the convex order literature stating that, for any $d \geq 2$ exchangeable random variables $X_1, \ldots, X_d$ and any convex function $\phi$, we have
\[
 \E\left[\phi\left(\frac{1}{d} \sum_{i=1}^d X_i\right)\right] \leq \E\left[\phi\left(\frac{1}{d - 1} \sum_{i=1}^{d - 1} X_i\right)\right],
\]
whenever the expectations exist \citep[Corollary 1.5.24]{muller2002comparison}. We are thus able to conclude by setting $X_i = \epsilon_i^2$ and $\phi(x) = \Phi\left(-(\ell / 2) \surd{x}\right)$ for all $x \geq 0$ given that this function is convex.
\end{proof}

\begin{proof}[Proof of \autoref{prop:scaling_const}]
 We prove that
 \[
  2\lambda^2 \, \E\left\{ \|\boldsymbol\epsilon\|^2 \, \Phi\left(-\lambda \,  \frac{\|\boldsymbol\epsilon\|}{2}\right) \right\} = 2\ell^2 \, \E\left\{ \frac{\|\boldsymbol\epsilon\|^2}{d} \, \Phi\left(-\ell \,  \frac{\|\boldsymbol\epsilon\|/\surd{d}}{2}\right) \right\} \rightarrow 2\ell^2 \, \Phi\left(- \frac{\ell}{2}\right).
 \]
 The convergence
 \[
  2 \, \E\left\{ \Phi\left(-\lambda \,  \frac{\|\boldsymbol\epsilon\|}{2}\right) \right\} =   2 \, \E\left\{ \Phi\left(-\ell \,  \frac{\|\boldsymbol\epsilon\|/\surd{d}}{2}\right) \right\}  \rightarrow 2 \, \Phi\left(- \frac{\ell}{2}\right)
 \]
 follows using similar arguments.

 By the  strong law of large numbers, we have that $\|\boldsymbol\epsilon\|^2 / d \rightarrow 1$ almost surely, and then
 \[
  \frac{\|\boldsymbol\epsilon\|^2}{d} \, \Phi\left(-\ell \,  \frac{\|\boldsymbol\epsilon\|/\surd{d}}{2}\right) \rightarrow \Phi\left(- \frac{\ell}{2}\right),
 \]
 almost surely. To prove that the expectation converges, we show that
 \[
  \frac{\|\boldsymbol\epsilon\|^2}{d} \, \Phi\left(-\ell \,  \frac{\|\boldsymbol\epsilon\|/\surd{d}}{2}\right)
 \]
 is uniformly integrable. To prove this, we show that
 \[
  \sup_d \E\left\{\left( \frac{\|\boldsymbol\epsilon\|^2}{d} \, \Phi\left(-\ell \,  \frac{\|\boldsymbol\epsilon\|/\surd{d}}{2}\right)\right)^2 \right\} < \infty.
 \]
 Using that $0 \leq \Phi \leq 1$ and that $\|\boldsymbol\epsilon\|^2$ has a Chi-square distribution with $d$ degrees of freedom,
 \begin{align*}
 \E\left\{\left( \frac{\|\boldsymbol\epsilon\|^2}{d} \, \Phi\left(-\ell \,  \frac{\|\boldsymbol\epsilon\|/\surd{d}}{2}\right)\right)^2 \right\} \leq \E\left\{\left( \frac{\|\boldsymbol\epsilon\|^2}{d} \, \right)^2 \right\} = \frac{2d + d^2}{d^2},
 \end{align*}
 which has a finite supremum. This concludes the proof that
 \[
  2\ell^2 \, \E\left\{ \frac{\|\boldsymbol\epsilon\|^2}{d} \, \Phi\left(-\ell \,  \frac{\|\boldsymbol\epsilon\|/\surd{d}}{2}\right) \right\} \rightarrow 2\ell^2 \, \Phi\left(- \frac{\ell}{2}\right).
 \]

 The function $2\ell^2 \, \Phi\left(- \frac{\ell}{2}\right)$ can be optimized numerically and is maximized by $\ell = \hat{\ell} := 2.38$.
\end{proof}

\bibliography{references}

\end{document}